\documentclass[fleqn,usenatbib]{mnras}

\usepackage[T1]{fontenc}
\usepackage{ae,aecompl}

\usepackage{times} 
\usepackage{tikz}
\usetikzlibrary{positioning}
\usepackage{xspace,ulem}

\usetikzlibrary{backgrounds}
\usetikzlibrary{arrows}
\usetikzlibrary{calc}
\usepackage{booktabs}
\usepackage{graphicx}	
\usepackage{amsmath}	
\usepackage{amssymb}	
\usepackage{times}
\usepackage{newtxtext}
\usepackage[varvw]{newtxmath}

\title[Pipeline for $f(R)$ cluster constraints]{\boldmath A general framework to test gravity using galaxy clusters V: \\ A self-consistent pipeline for unbiased constraints of $f(R)$ gravity}

\author[M.~A.~Mitchell et al.]{
Myles A.~Mitchell,\thanks{E-mail: m.a.mitchell@durham.ac.uk}
Christian Arnold
and Baojiu Li
\\
Institute for Computational Cosmology, Department of Physics, Durham University, South Road, Durham DH1 3LE, UK
}

\date{Accepted XXX. Received YYY; in original form ZZZ}

\pubyear{2021}

\begin{document}
\label{firstpage}
\pagerange{\pageref{firstpage}--\pageref{lastpage}}
\maketitle

\begin{abstract}
We present a Markov chain Monte Carlo pipeline that can be used for robust and unbiased constraints of $f(R)$ gravity using galaxy cluster number counts. This pipeline makes use of a detailed modelling of the halo mass function in $f(R)$ gravity, which is based on the spherical collapse model and calibrated by simulations, and fully accounts for the effects of the fifth force on the dynamical mass, the halo concentration and the observable-mass scaling relations. Using a set of mock cluster catalogues observed through the thermal Sunyaev-Zel'dovich effect, we demonstrate that this pipeline, which constrains the present-day background scalar field $f_{R0}$, performs very well for both $\Lambda$CDM and $f(R)$ fiducial cosmologies. We find that using an incomplete treatment of the scaling relation, which could deviate from the usual power-law behaviour in $f(R)$ gravity, can lead to imprecise and biased constraints. We also find that various degeneracies between the modified gravity, cosmological and scaling relation parameters can significantly affect the constraints, and show how this can be rectified by using tighter priors and better knowledge of the cosmological and scaling relation parameters. Our pipeline can be easily extended to other modified gravity models, to test gravity on large scales using galaxy cluster catalogues from ongoing and upcoming surveys.
\end{abstract}

\begin{keywords}
cosmology: theory, dark energy -- galaxies: clusters: general -- methods: numerical
\end{keywords}



\section{Introduction}
\label{sec:introduction}

Galaxy clusters are the largest virialised objects in the Universe to have been observed, and are believed to trace the highest peaks of the primordial density fluctuations. Their abundance is highly sensitive to the values of a number of cosmological parameters, including the matter density parameter $\Omega_{\rm M}$ and the linear density fluctuation $\sigma_8$, which both affect the formation of large-scale structure. They are also sensitive to the strength of gravity on large scales, and can therefore be used to constrain modified gravity (MG) theories \citep[e.g.,][]{Koyama:2015vza} which have been proposed in order to explain the late-time accelerated cosmic expansion. Various ongoing and upcoming astronomical surveys are generating vast cluster catalogues using all means of detection, including the clustering of galaxies \citep[e.g.,][]{ukidss,lsst,euclid,desi}, distortions of the cosmic microwave background (CMB) by the Sunyaev-Zel'dovich (SZ) effect \citep[e.g.,][]{act,Planck_SZ_cluster,Abazajian:2016yjj,Ade:2018sbj}, and X-ray emission from the hot intra-cluster gas \citep[][]{chandra,xmm-newton,erosita}. These will be many times larger than previous catalogues, and will significantly advance our understanding of gravity at the largest scales.

Before we can use this data, it is necessary to prepare robust theoretical predictions that can be combined with the observations to make constraints. In particular, special care should be given to potential sources of bias. For example, many MG theories predict a strengthened gravitational force in certain regimes \citep[e.g.,][]{DVALI2000208,Hu:2007nk}. In addition to enhancing the abundance of clusters, this can affect internal properties including the density profile and the temperature of the intra-cluster gas. A consequence of this is that observable-mass scaling relations $Y(M)$, which can be used to relate the cluster mass $M$ to some observable $Y$, can deviate from General Relativity (GR) predictions \citep[see, e.g.,][]{He:2015mva}. Scaling relations are a vital ingredient for cluster cosmology: for example, they are used to relate the observational mass function, with the form ${\rm d}n/{\rm d}Y$, to the theoretical mass function ${\rm d}n/{\rm d}M$, and to infer cluster observables in mock catalogues. They have therefore been widely studied both theoretically with numerical simulations \citep[e.g.,][]{Fabjan:2011,Truong:2016egq} and with observations \citep[e.g.,][]{Ade:2013lmv}. Understanding how they are affected by a strengthened gravity is crucial in order to prevent biased estimates of the cluster mass.

Great advances have been made in recent years in the development of subgrid models for baryonic processes including star formation, cooling and stellar and black hole feedback \citep[e.g.,][]{Schaye:2014tpa,2017MNRAS.465.3291W,Pillepich:2017jle}. By including these in cosmological simulations, it has become possible to simulate populations of galaxies whose gaseous and stellar properties closely match real observations \citep[e.g.,][]{2014Natur.509..177V}. It is important to include these `full physics' models in MG simulations \citep[e.g.,][]{Arnold:2019vpg,Hernandez-Aguayo:2020kgq} in order to understand the full impact of MG forces on the thermal properties of clusters. This can pave the way for a complete treatment of the scaling relations in constraint pipelines, which can then be used to infer unbiased large-scale constraints of gravity.

The $f(R)$ gravity model \citep[e.g.,][]{10.1093/mnras/150.1.1,Sotiriou:2008rp,DeFelice:2010aj} is a particularly popular and well-studied MG model that can provide an alternative explanation for the late-time accelerated expansion, but, more importantly, can be employed to study the viability of modifications to GR. This includes an additional `fifth force' which enhances the total strength of gravity. The fifth force leaves numerous observational signatures in large-scale structure, and the theory can be tested using a variety of probes, including cluster number counts \citep[e.g.,][]{PhysRevD.92.044009,Liu:2016xes,Peirone:2016wca}, redshift-space distortions \citep[e.g.,][]{Bose:2017dtl,2018NatAs...2..967H,Hernandez-Aguayo:2018oxg}, the cluster gas mass fraction \citep[e.g.,][]{Li:2015rva}, the clustering of clusters \citep{Arnalte-Mur:2016alq}, the cluster SZ profile \citep{deMartino:2016xso}, the SZ angular power spectrum \citep{Mitchell:2020fnj} and weak lensing by voids \citep{Cautun:2017tkc}. As described above, the fifth force affects the observable-mass scaling relations of clusters, which can themselves be used to probe gravity \citep[see, e.g.,][]{Hammami:2016npf,DelPopolo:2019oxn}. The fifth force also causes the dynamical mass of clusters to become enhanced with respect to the lensing mass \citep{arnold:2014}. This has been used to probe the theory by comparing weak lensing measurements of clusters with thermal observations \citep[e.g.,][]{Terukina:2013eqa,Wilcox:2015kna}.
 
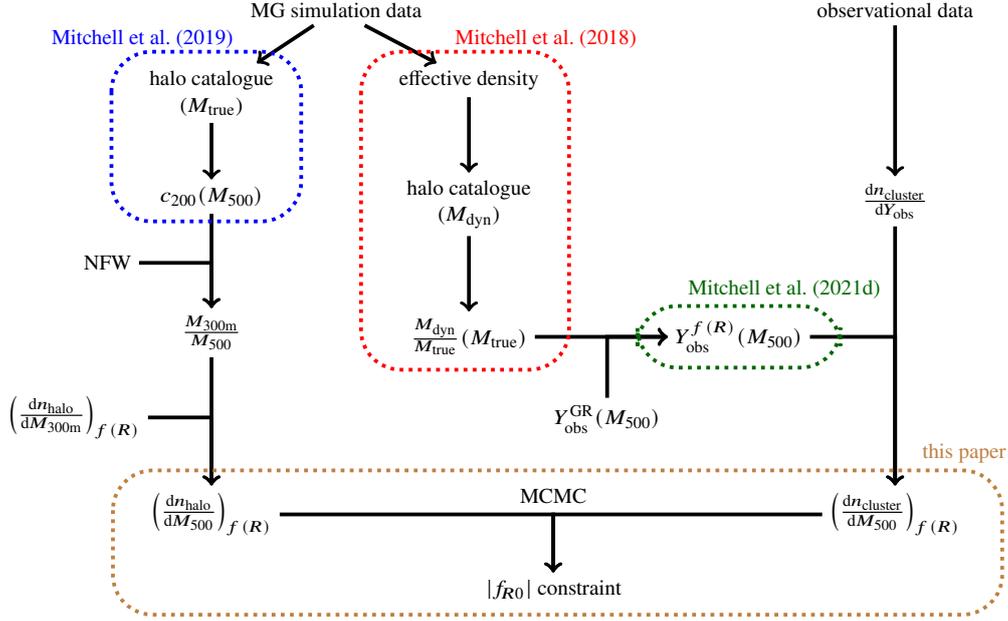
\begin{figure*}
\centering
\begin{tikzpicture}
\tikzstyle{myarrow}=[line width=0.5mm,draw=black,-triangle 45,postaction={draw, line width=0.5mm, shorten >=4mm, -}]

\node    (simulations)    {MG simulation data};
\node    (cat_true)    [below left = 0.5cm and -0.5cm of simulations]   [align=center]{halo catalogue \\ ($M_{\rm true}$)};
\node    (c_m)    [below = 0.75cm of cat_true] [align=center]   {$c_{\rm 200}(M_{\rm 500})$};
\node   (m300_m500)    [below = 1.25cm of c_m] [align=center]    {$\frac{M_{\rm 300m}}{M_{500}}$};
\node    (nfw)    [below left = 0.65cm and 1.0cm of c_m]  [anchor=west]   {NFW};
\node    (hmf_m300)    [below left = 2.7cm and 2.0cm of c_m]  [anchor=west]   {$\left( \frac{{\rm d} n_{\rm halo}}{{\rm d}M_{\rm 300m}} \right)_{f(R)}$};
\node    (hmf_th)    [below = 1.7cm of m300_m500]  [align=center]   {$\left( \frac{{\rm d} n_{\rm halo}}{{\rm d} M_{500}} \right)_{f(R)}$};
\node    (rho_eff)    [below right = 0.5cm and -0.5cm of simulations]   [align=center]  {effective density};
\node    (cat_mdyn)    [below = 1.0cm of rho_eff]    [align=center] [align=center]{halo catalogue \\ ($M_{\rm dyn}$)};
\node    (mdyn_mtrue)    [below = 1.0cm of cat_mdyn]  [align=center]   {$\frac{M_{\rm dyn}}{M_{\rm true}}(M_{\rm true})$};
\node    (observations)    [right = 5cm of simulations]    [align=center] {observational data};
\node    (n_Y)    [below = 2.0cm of observations]   [align=center]  {$\frac{{\rm d} n_{\rm cluster}}{{\rm d}Y_{\rm{obs}}}$};
\node    (hmf_obs)  at (hmf_th -| n_Y)  [ align=center]   {$\left(\frac{{\rm d} n_{\rm cluster}}{{\rm d} M_{\rm 500}}\right)_{f(R)}$};
\node    (scaling_relation)    [right = 1.75cm of mdyn_mtrue]    [align=center]{$Y_{\rm{obs}}^{f(R)}(M_{500})$};
\node    (scaling_relation_lcdm)    [below left= 0.5cm and -0cm of scaling_relation]    [align=center]{$Y_{\rm{obs}}^{\rm GR}(M_{500})$};

\node    (mcmc)    at ($(hmf_th)!0.5!(hmf_obs)+(0.0,0.25)$)   [align=center]  {MCMC};
\node    (constraint)  at ($(hmf_th)!0.5!(hmf_obs)+(0.0,-1.0)$) [align=center]  {$|f_{R0}|$ constraint};

\draw[->, line width=0.5mm] (simulations) -- (cat_true);
\draw[->, line width=0.5mm] (cat_true) -- (c_m);
\draw[->, line width=0.5mm] (c_m) -- (m300_m500);
\draw[->, line width=0.5mm, to path={-| (\tikztotarget)}] (nfw) edge (m300_m500);
\draw[->, line width=0.5mm] (m300_m500) -- (hmf_th);
\draw[->, line width=0.5mm, to path={-| (\tikztotarget)}] (hmf_m300) edge (hmf_th);
\draw[->, line width=0.5mm] (simulations) -- (rho_eff);
\draw[->, line width=0.5mm] (rho_eff) -- (cat_mdyn);
\draw[->, line width=0.5mm] (cat_mdyn) -- (mdyn_mtrue);

\draw[->, line width=0.5mm] (observations) -- (n_Y);
\draw[->, line width=0.5mm] (n_Y) -- (hmf_obs);
\draw[->, line width=0.5mm, to path={-| (\tikztotarget)}] (scaling_relation) edge (hmf_obs);
\draw[->, line width=0.5mm] (mdyn_mtrue) -- (scaling_relation);
\draw[->, line width=0.5mm, to path={|- (\tikztotarget)}] (scaling_relation_lcdm) edge (scaling_relation);

\draw[->, line width=0.5mm, to path={-| (\tikztotarget)}] (hmf_th) edge (constraint);
\draw[->, line width=0.5mm, to path={-| (\tikztotarget)}] (hmf_obs) edge (constraint);


\draw [line width=0.5mm,dotted, red, rounded corners=15pt]     ($(rho_eff.north west)+(-0.4,0.15)$) rectangle ($(mdyn_mtrue.south east)+(0.45,-0.1)$);
\node [above right = 0.10cm and -1.3cm of rho_eff] {\small{\color{red}\hypersetup{citecolor=red}\cite{Mitchell:2018qrg}}}; 
\draw[line width=0.5mm,dotted, blue, rounded corners=15pt]   ($(cat_true.north west)+(-0.4,0.15)$) rectangle ($(c_m.south east)+(0.4,-0.1)$);
\node [above left = 0.10cm and -1.3cm of cat_true] {\small{\color{blue}\hypersetup{citecolor=blue}\cite{Mitchell:2019qke}}}; 
\draw[line width=0.5mm,dotted, brown, rounded corners=15pt]     ($(hmf_th.north west)+(-0.4,0.2)$) rectangle ($(constraint.south east -| hmf_obs.south east) +(0.45,-0.1)$);
\node [above right = 0.20cm and -0.7cm of hmf_obs] {\small{\color{brown}this paper}}; 
\draw[line width=0.5mm,dotted, black!60!green, rounded corners=15pt]     ($(scaling_relation.north west)+(-0.4,0.1)$) rectangle ($(scaling_relation.south east) +(0.4,-0.1)$);
\node [above right = 0.10cm and -1.7cm of scaling_relation] {\small{\color{black!60!green}\hypersetup{citecolor=black!60!green}\cite{Mitchell:2020aep}}}; 
\end{tikzpicture}
\caption{[{\it Colour Online}] Flow chart outlining our general framework for constraining the present-day background scalar field, $f_{R0}$, of $f(R)$ gravity using the cluster abundance. We use our model for the halo concentration, calibrated in \citet{Mitchell:2019qke} (\textit{blue dotted box}), to convert the theoretical prediction of the HMF in $f(R)$ gravity, which is based on the model by \citet{Cataneo:2016iav}, from mass definition $M_{\rm 300m}$ to $M_{500}$. We model the $f(R)$ observable-mass scaling relation (\textit{green dotted box}) by rescaling a GR power-law relation using our model for the ratio of the dynamical mass to the true mass, calibrated in \citet{Mitchell:2018qrg} (\textit{red dotted box}); this method has been tested and verified using $f(R)$ simulations that include full baryonic physics \citep{Mitchell:2020aep}. The scaling relation is used to relate the observational form of the mass function, ${\rm d}n/{\rm d}Y_{\rm obs}$, to the theoretical form ${\rm d}n/{\rm d}M_{500}$. In this work, we test our MCMC pipeline for constraining $f_{R0}$ (\textit{brown dotted box}) using the theoretical HMF and data from mock cluster catalogues.}
\label{fig:flow_chart}
\end{figure*}

This paper is part of a series of works which are aimed at developing a general framework for unbiased cluster constraints of gravity. So far, we have modelled the effects of the $f(R)$ gravity fifth force on the cluster dynamical mass and the halo concentration using a suite of dark-matter-only simulations \citep{Mitchell:2018qrg,Mitchell:2019qke}, and we recently used the first simulations that simultaneously incorporate full physics and $f(R)$ gravity to study the effect on observable-mass scaling relations \citep{Mitchell:2020aep}. Our framework is designed to be extended beyond $f(R)$ gravity to other MG theories; indeed, we recently modelled cluster properties and the halo mass function (HMF) in the normal-branch Dvali-Gabadadze-Porrati model (nDGP) \citep{Mitchell:2021aex}. 

Fig.~\ref{fig:flow_chart} gives a broad overview of our proposed framework for $f(R)$ constraints using cluster number counts: our model for the enhancement of the concentration (blue dotted box) can be used for conversions between cluster mass definitions, which is required if, for example, the theoretical predictions and observations use different spherical overdensities; our model for the dynamical mass enhancement (red dotted box) can be used to predict the $f(R)$ scaling relation (green dotted box) given a GR counterpart relation, and this can be used to relate ${\rm d}n/{\rm d}Y$ to ${\rm d}n/{\rm d}M$; finally, the observations and theoretical predictions are combined to constrain the present-day background scalar field using Markov-chain Monte Carlo (MCMC) sampling (brown dotted box). In this work, we test this pipeline using mock cluster catalogues generated for both GR and $f(R)$ fiducial cosmologies. In doing so, we can assess the importance of using a complete modelling of the scaling relation, which behaves as a broken power law in $f(R)$ gravity. We also consider other potential sources of bias which can arise from degeneracies among model parameters.

This paper is arranged as follows: in Sec.~\ref{sec:background}, we provide an overview of the $f(R)$ gravity theory and our models for the effects of the fifth force on the cluster properties; in Sec.~\ref{sec:methods}, we describe our MCMC constraint pipeline, including the calculation of the log-likelihood and the generation of the mocks; in Sec.~\ref{sec:results}, we present 
constraints using the GR and $f(R)$ mocks; then, in Sec.~\ref{sec:bias}, we highlight potential sources of bias in our pipeline; finally, we summarise our main findings in Sec.~\ref{sec:conclusions}.

Throughout this work, we denote background quantities with overbars ($\bar{x}$), Greek indices can take values 0, 1, 2 and 3, and we use the unit convention $c=1$ for the speed of light. This paper will use different spherical mass definitions for dark matter haloes, based on the following rule: $M_\Delta$ means the mass enclosed by halo radius $R_\Delta$, within which the mean matter density is $\Delta$ times the critical density at the halo redshift. We will mostly use 3 values of $\Delta$: $200, 500$ and $300\Omega_{\rm M}(z)$, with $\Omega_{\rm M}(z)$ the matter density parameter at redshift $z$; for the first two the mass is respectively written as $M_{200}$ and $M_{500}$, while for the last the notation is $M_{300{\rm m}}$.

\section{Background}
\label{sec:background}

In Sec.~\ref{sec:background:theory}, we describe the underlying theory of $f(R)$ gravity. Then, in Sec.~\ref{sec:background:clusters}, we outline the effects of $f(R)$ gravity on the properties of galaxy clusters.

\subsection{Theory}
\label{sec:background:theory}

In the $f(R)$ gravity model, the gravitational action is given by:
\begin{equation}
    S=\int {\rm d}^4x\sqrt{-g}\left[\frac{R+f(R)}{16\pi G}+\mathcal{L}_{\rm M}\right],
\label{eq:action}
\end{equation}
where $g$ is the determinant of the metric tensor $g_{\alpha\beta}$, $R$ is the Ricci scalar curvature, $G$ is Newton's gravitational constant and $\mathcal{L}_{\rm M}$ is the Lagrangian matter density. The extra nonlinear curvature-dependent function $f(R)$ represents a modification to the Einstein-Hilbert action of GR. This leads to the modified Einstein field equations:
\begin{equation}
    G_{\alpha \beta} + X_{\alpha \beta} = 8\pi GT_{\alpha \beta},
\label{eq:modified_field_equations}
\end{equation}
where $G_{\alpha\beta}$ is the Einstein tensor and $T_{\alpha\beta}$ is the stress-energy tensor. The extra tensor $X_{\alpha\beta}$ encapsulates the modifications to GR, and is given by:
\begin{equation}
    X_{\alpha \beta} = f_RR_{\alpha \beta} - \left(\frac{f}{2}-\Box f_R\right)g_{\alpha \beta} - \nabla_{\alpha}\nabla_{\beta}f_R,
\label{eq:GR_modification}
\end{equation}
where $R_{\alpha\beta}$ is the Ricci curvature tensor, $\Box \equiv \nabla_\alpha \nabla^\alpha$ is the d'Alembert operator (using Einstein's summation convention) and $\nabla_{\alpha}$ represents the covariant derivative with respect to coordinate $\alpha \in \{0,1,2,3\}$ associated with the metric. The quantity $f_R\equiv{\rm d}f(R)/{\rm d}R$ represents the extra scalar degree of freedom of the theory, and is referred to as the scalar (or `scalaron') field. This mediates a fifth force which, when able to act, enhances the total strength of gravity by up to a factor of $4/3$. The fifth force can only act on scales smaller than the Compton wavelength:
\begin{equation}
    \lambda_{\rm C} = a^{-1}\left(3\frac{{\rm d}f_R}{{\rm d}R}\right)^{\frac{1}{2}},
\label{eq:compton_wavelength}
\end{equation}
where $a$ is the cosmic scale factor. 

The $f(R)$ model features a screening mechanism which can help to ensure consistency with Solar System tests \citep{Will:2014kxa}. This is achieved by giving the scalaron an environment-dependent effective mass which becomes larger in dense regions, suppressing its gravitational interaction. Consequently, the fifth force can only act in sufficiently low-density regions which can include, for example, cosmic voids, low-mass haloes and the outer regions of galaxy clusters, where the gravitational potential well is not too deep. The masking of the fifth force in $f(R)$ gravity, which is achieved through the nonlinear total interaction potential that appears in the Lagrangian of the scalaron, is an example of the chameleon screening mechanism \citep[e.g.,][]{Khoury:2003aq,Khoury:2003rn,Mota:2006fz}.

The \citet{Hu:2007nk} model of $f(R)$ gravity assumes the following prescription for the function $f(R)$:
\begin{equation}
    f(R) = -m^2\frac{c_1\left(-R/m^2\right)^n}{c_2\left(-R/m^2\right)^n+1},
\label{eq:hu_sawicki}
\end{equation}
where $n$, $c_1$ and $c_2$ are the free parameters of the model. The quantity $m^2$ is equivalent to $8\pi G\bar{\rho}_{\rm M,0}/3=H_0^2\Omega_{\rm M}$, where $\bar{\rho}_{\rm M,0}$ is the present-day mean matter density and $H_0$ is the present-day Hubble parameter. By choosing $c_1/c_2=6\Omega_{\Lambda}/\Omega_{\rm M}$ and assuming the inequality $-\bar{R}\gg m^2$ for the background curvature, it can be shown that $f(R)$ behaves as a cosmological constant in background cosmology \citep{Hu:2007nk}. 

Assuming the above inequality, we obtain the following approximation for the background scalar field:
\begin{equation}
\bar{f_R} \approx -n\frac{c_1}{c_2^2}\left(\frac{m^2}{-\bar{R}}\right)^{n+1}.
\label{eq:scalar_field}
\end{equation}
The background curvature is given by:
\begin{equation}
-\bar{R} = 3m^2\left(a^{-3}+4\frac{\Omega_\Lambda}{\Omega_{\rm M}}\right),
\label{eq:R}
\end{equation}
where $\Omega_{\Lambda}=1-\Omega_{\rm M}$. The parameter combination $c_1/c_2^2$ can then be rewritten as:
\begin{equation}
\frac{c_1}{c_2^2} = -\frac{1}{n}\left[3\left(1+4\frac{\Omega_{\Lambda}}{\Omega_{\rm M}}\right)\right]^{n+1}f_{R0},
\label{eq:c1/c22}
\end{equation}
where $f_{R0}$ is the present-day value of the background scalar field (we will omit the over-bar for this quantity throughout this work). From Eq.~(\ref{eq:R}), we see that the inequality $-\bar{R}\gg m^2$ is valid for realistic values of $\Omega_{\rm M}$. Using this approximation, we have reformulated the original 3-parameter model into a form that has just two free parameters: $n$ and $f_{R0}$. In this work, we will set $n=1$, which is a common choice in literature. Therefore, $f_{R0}$ is the parameter that we aim to probe with our $f(R)$ gravity constraint pipeline. From Eqs.~(\ref{eq:scalar_field}) and (\ref{eq:R}), we see that the background scalar field has a greater amplitude at later times, therefore $|f_{R0}|$ represents the highest amplitude in cosmic history. In $f(R)$ models with a higher $|f_{R0}|$, the fifth force can be felt by haloes with a greater mass. These models therefore represent a greater departure from GR. In this work, we will use the naming convention F6.5, F6, F5.5, ..., F4 when referring to models with $|f_{R0}|=10^{-6.5},10^{-6},10^{-5.5},...,10^{-4}$ (in order from weakest to strongest).

In the following subsections, we will summarise the effects of the $f(R)$ fifth force on the properties of haloes.

\subsection{\boldmath Galaxy clusters in \texorpdfstring{$f(R)$}{f(R)} gravity}
\label{sec:background:clusters}

In this section, we summarise the main effects of $f(R)$ gravity on the properties of galaxy clusters. We present our models for the enhancements of the dynamical mass and halo concentration in Secs.~\ref{sec:background:dynamical_mass} and \ref{sec:background:concentration}. Then, in Sec.~\ref{sec:background:scaling_relations}, we show how our model for the dynamical mass enhancement can be used to map between scaling relations in $f(R)$ gravity and GR. Finally, in Sec.~\ref{sec:background:hmf}, we outline the modelling by \citet{Cataneo:2016iav} for the $f(R)$ enhancement of the HMF.

\subsubsection{Dynamical mass enhancement and its scatters}
\label{sec:background:dynamical_mass}

Throughout this work, we will refer to two mass definitions that are applicable to, but not exclusive to, haloes. We define the `dynamical' mass as the mass that is felt by a nearby massive test particle: this can be inferred from observations related to the gravitational potential of the halo, including the gas temperature and the virial velocities of galaxies \citep[e.g.,][]{Farahi:2016xux,Evrard:2007py}. Meanwhile, we define the `true' mass as the intrinsic mass: in simulations this is equivalent to the summed mass of the halo particles, while in observations this would be inferred using lensing data (photons hardly feel the fifth force for viable $f(R)$ models). When the fifth force is able to act, the dynamical mass of a halo is enhanced relative to the true mass: $M_{\rm true}^{f(R)}\leq M_{\rm dyn}^{f(R)}\leq (4/3)M_{\rm true}^{f(R)}$. Meanwhile, the two masses are expected to be equal in GR: $M_{\rm true}^{\rm GR}=M_{\rm dyn}^{\rm GR}=M^{\rm GR}$.

In \citet{Mitchell:2018qrg}, we used a suite of dark-matter-only simulations, which span a wide range of resolutions and box sizes, to calibrate a general formula for the ratio $\mathcal{R}$ of the dynamical mass to the true mass:
\begin{equation}
\mathcal{R} = \frac{M^{f(R)}_{\rm dyn}}{M^{f(R)}_{\rm true}} = \frac{7}{6}-\frac{1}{6}\tanh\left(p_1\left[\log_{10}\left(M^{f(R)}_{\rm true}M_{\odot}^{-1}h\right)-p_2\right]\right),
\label{eq:mdyn_enhancement}
\end{equation}
where $h=H_0/(100~{\rm kms}^{-1}{\rm Mpc}^{-1})$, and $p_1$ and $p_2$ are the model parameters. We found that $p_1$ is approximately constant, with best-fit value $2.21\pm0.01$, while the parameter $p_2$ closely follows the following physically motivated linear relation:
\begin{equation}
    p_2=(1.503\pm0.006)\log_{10}\left(\frac{|\bar{f}_R(z)|}{1+z}\right)+(21.64\pm0.03).
\label{eq:p2}
\end{equation}
Physically, $p_2$ represents the logarithmic mass above which haloes are mostly screened and below which haloes are mostly unscreened. This model can accurately reproduce the dynamical mass enhancement for haloes in the mass range $10^{11}h^{-1}M_{\odot}\lesssim M_{500}\lesssim 10^{15}h^{-1}M_{\odot}$ with redshifts $0\leq z\leq1$, for models with present-day field strengths $10^{-6.5}\leq |f_{R0}|\leq 10^{-4}$.

\begin{figure}
\centering
\includegraphics[width=\columnwidth]{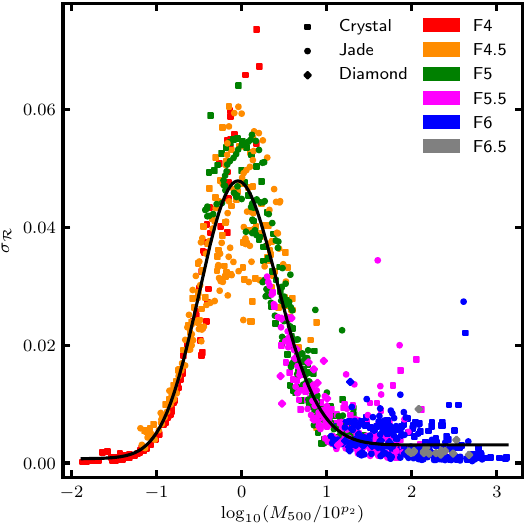}
\caption{[{\it Colour Online}] Root-mean-square scatter in the dynamical mass enhancement as a function of the rescaled halo mass $\log_{10}(M_{500}/10^{p_2})$, where $p_2$ is given by Eq.~(\ref{eq:p2}). The data points, which correspond to mass bins spanning $10^{11}h^{-1}M_{\odot}\lesssim M_{500}\lesssim 10^{15}h^{-1}M_{\odot}$, have been generated using the dark-matter-only simulations Crystal (\textit{squares}), Jade (\textit{circles}) and Diamond (\textit{diamonds}), which are described in \citet{Mitchell:2018qrg}. The data spans redshifts $0\leq z\leq1$ and includes present-day scalar field amplitudes $|f_{R0}|=10^{-6.5}$ (\textit{grey}), $10^{-6}$ (\textit{blue}), $10^{-5.5}$ (\textit{magenta}), $10^{-5}$ (\textit{green}), $10^{-4.5}$ (\textit{orange}) and $10^{-4}$ (\textit{red}). The solid line represents our best-fit model, which is given by Eq.~(\ref{eq:scatter_model}).}
\label{fig:rms_scatter}
\end{figure}

For this work, we have again used the data from \citet{Mitchell:2018qrg} to model the root-mean-square scatter of the dynamical mass enhancement, $\sigma_{\mathcal{R}}$. Our model is shown by the solid line in Fig.~\ref{fig:rms_scatter} (we provide a detailed description of this model and our fitting procedure in Appendix \ref{sec:appendix:mdyn_scatter}). The rescaled mass, $\log_{10}(M_{500}M_{\odot}^{-1}h)-p_2\equiv\log_{10}(M_{500}/10^{p_2})$, is expected to take positive values for haloes that are screened and negative values for haloes that are unscreened. The scatter peaks for haloes that are partially screened, with $\log_{10}(M_{500}M_{\odot}^{-1}h)\sim p_2$, whereas it falls to roughly zero for lower and higher masses. Physically, this makes sense: at sufficiently high masses where \textit{all} haloes are screened and have $\mathcal{R}\approx1$, it follows that the scatter $\sigma_{\mathcal{R}}$ is very small, and a similar argument can be applied for haloes deep in the unscreened regime. Between these two regimes, the physics is more complicated, giving rise to greater dispersion in the chameleon screening; for example, haloes which do not have a high enough mass to be self-screened can still be environmentally screened by nearby massive haloes.

\subsubsection{Halo concentration}
\label{sec:background:concentration}

The halo concentration, $c$, is a parameter of the universal Navarro-Frenk-White (NFW) density profile of dark matter haloes \citep{NFW}. For a given halo mass definition, such as $M_{200}$ or $M_{500}$, if the mass (or radius) of a halo is known, then the concentration is the only parameter required to describe the halo density profile. In \citet{Mitchell:2019qke}, we used an extensive suite of dark-matter-only simulations to study the effects of the fifth force on the concentration in $f(R)$ gravity. There are a range of behaviours, depending on the level of screening: the concentration of recently-unscreened haloes can be enhanced by up to $\sim40\%$ as particles concentrate at the inner regions; however, for haloes with lower masses that have been unscreened for longer, the enhancement of the concentration drops over time as the halo particles whose kinetic energy is boosted gradually migrate away from, or manage to stay away from, the halo centre. And at higher masses, where haloes are only unscreened at the outermost regions, the concentration can be suppressed by up to $\sim5\%$. We modelled this behaviour using the rescaled logarithmic mass $x=\log_{10}(M_{500}/10^{p_2})$ defined in Sec.~\ref{sec:background:dynamical_mass}, and found excellent agreement with the following formula:
\begin{equation}
\begin{split}
\log_{10}\left|\frac{c}{c_{\rm GR}}\right|_{200} = &\frac{1}{2}\left(\frac{\lambda}{\omega_{\rm s}}\phi(x')\left[1+\rm{erf}\left(\frac{\alpha x'}{\sqrt[]{2}}\right)\right]+\gamma\right)\\
&\times(1-\tanh\left(\omega_{\rm t}\left[x+\xi_{\rm t}\right]\right)),
\end{split}
\label{eq:c_model}
\end{equation}
where $c_{200}$ denotes the concentration of haloes with mass definition $M_{200}$. This is a product of a skewed normal distribution and a tanh formula, where $x'=(x-\xi_{\rm s})/\omega_{\rm s}$, $\phi(x')$ is the normal distribution and $\rm{erf}(\alpha x'/\sqrt{2})$ is the error function. The parameters have best-fit values $\lambda=0.55$, $\omega_{\rm s}=1.7$, $\xi_{\rm s}=-0.27$, $\alpha=-6.5$, $\gamma=-0.07$, $\omega_{\rm t}=1.3$ and $\xi_{\rm t}=0.1$.

\subsubsection{Observable-mass scaling relations}
\label{sec:background:scaling_relations}

The thermal properties of the intra-cluster gas are intrinsically related to the gravitational potential of the halo \citep[e.g.,][]{1986MNRAS.222..323K,Voit:2004ah}. This is because, during cluster formation, the initial potential energy of nearby gas gets converted into thermal energy through shock-heating as it is accreted by the halo. As a result, various cluster observables -- including the gas temperature $T_{\rm gas}$, the Compton $Y$-parameter of the SZ effect ($Y_{\rm SZ}$) and the X-ray analogue of the $Y$-parameter ($Y_{\rm X}$) -- have one-to-one (for an ideal situation) mappings with the cluster mass. As discussed in Sec.~\ref{sec:introduction}, these scaling relations are a vital ingredient for cluster cosmology.

In \citet{Mitchell:2020aep}, we used cosmological simulations which incorporate full baryonic physics to verify a set of mappings, originally proposed by \citet{He:2015mva}, between the $f(R)$ scaling relations and their GR power-law counterparts. The mapping for the $Y_{\rm SZ}$ observable is given by:
\begin{equation}
    \frac{M_{\rm dyn}^{f(R)}}{M_{\rm true}^{f(R)}}Y_{\rm SZ}^{f(R)}\left(M_{\rm dyn}^{f(R)}\right) \approx Y_{\rm SZ}^{\rm GR}\left(M^{\rm GR}=M_{\rm dyn}^{f(R)}\right).
    \label{eq:ysz_mapping}
\end{equation}
This says that the $Y_{\rm SZ}$ parameter of an $f(R)$ halo with dynamical mass $M_{\rm dyn}^{f(R)}$ differs by a factor of $M_{\rm dyn}^{f(R)}/M_{\rm true}^{f(R)}$ compared to a GR halo with the same mass $M^{\rm GR}=M_{\rm dyn}^{f(R)}$. The temperatures of these two haloes are the same, since they have the same total gravitational potential (including the fifth force contribution): 
\begin{equation}
    T^{f(R)}_{\rm gas}\left(M^{f(R)}_{\rm dyn}\right) = T^{\rm GR}_{\rm gas}\left(M^{\rm GR}=M^{f(R)}_{\rm dyn}\right).
    \label{eq:temp_equiv_eff}
\end{equation}
However, the gas density is higher in the GR halo compared to the $f(R)$ halo. This is because clusters form from matter found in an initially large region, such that the baryonic mass (mainly in the form of hot intracluster gas) and the total mass will follow the ratio between the cosmic baryonic and matter densities \citep[e.g.,][]{1993Natur.366..429W}; because the GR halo above has a higher intrinsic (true) mass than the F5 halo, it then follows that it also has a larger gas density. This gives rise to the $M_{\rm dyn}^{f(R)}/M_{\rm true}^{f(R)}$ factor in Eq.~(\ref{eq:ysz_mapping}), which can be predicted using Eq.~(\ref{eq:mdyn_enhancement}). The same mapping is predicted for the $Y_{\rm X}$ parameter, and a different mapping works for the cluster X-ray luminosity which will not be shown here. We showed that these mappings hold for halo masses $M_{500}\gtrsim10^{13.5}M_{\odot}$.

We also tested the following mapping for haloes in $f(R)$ gravity and GR that have the same true mass, $M_{\rm true}^{f(R)}=M^{\rm GR}$:
\begin{equation}
    Y_{\rm SZ}^{f(R)}\left(M_{\rm true}^{f(R)}\right) \approx \frac{M_{\rm dyn}^{f(R)}}{M_{\rm true}^{f(R)}}Y_{\rm SZ}^{\rm GR}\left(M^{\rm GR}=M_{\rm true}^{f(R)}\right).
    \label{eq:ysz_mapping_true}
\end{equation}
In this case, the total gravitational potential of the $f(R)$ haloes is enhanced by a factor of $M_{\rm dyn}^{f(R)}/M_{\rm true}^{f(R)}$ compared to the GR haloes. The temperature is then enhanced by the same factor: 
\begin{equation}
    T^{f(R)}_{\rm gas}\left(M^{f(R)}_{\rm true}\right) = \frac{M_{\rm dyn}^{f(R)}}{M_{\rm true}^{f(R)}}T^{\rm GR}_{\rm gas}\left(M^{\rm GR}=M^{f(R)}_{\rm true}\right).
    \label{eq:temp_equiv_true}
\end{equation}
This gives rise to the $M_{\rm dyn}^{f(R)}/M_{\rm true}^{f(R)}$ factor in Eq.~(\ref{eq:ysz_mapping_true}), and the same mapping is predicted for the $Y_{\rm X}$ parameter as well. We again showed that this mapping holds for halo masses $M_{500}\gtrsim10^{13.5}M_{\odot}$.

\subsubsection{Halo mass function}
\label{sec:background:hmf}

In this section, we will outline the \citet{Cataneo:2016iav} model for the $f(R)$ enhancement of the HMF, which we have adopted for our constraint pipeline. This is computed using the \citet{Sheth:1999mn} prescription of the HMF:
\begin{equation}
    n_{\rm ST} \equiv \frac{{\rm d}n}{{\rm d}\ln M} = \frac{\bar{\rho}_{\rm M}}{M}\frac{{\rm d}\ln\nu}{{\rm d}\ln M}\nu f(\nu),
    \label{eq:st_hmf}
\end{equation}
where the multiplicity function $\nu f(\nu)$ is given by:
\begin{equation}
    \nu f(\nu) = A\sqrt{\frac{2}{\pi}a\nu^2}\left[1+(a\nu^2)^{-p}\exp\left(-\frac{a\nu^2}{2}\right)\right].
    \label{eq:multiplicity}
\end{equation}
For the parameters $A$, $a$ and $p$, \citet{Cataneo:2016iav} used the fits by \citet{Despali:2015yla}, which extend the \citet{Sheth:1999mn} HMF to be a function of generic halo overdensity $\Delta$. For the latter, \citet{Cataneo:2016iav} used value $300\Omega_{\rm M}(z)$ (i.e., here the halo mass $M$ is $M_{300{\rm m}}$). The peak height $\nu$ is given by:
\begin{equation}
    \nu = \frac{\delta_{\rm c}}{\sigma(M,z)},
    \label{eq:nu}
\end{equation}
where $\delta_{\rm c}$ is the linearly extrapolated threshold density for spherical collapse and $\sigma(M,z)$ is the linear root-mean-square fluctuation of the matter density within spheres of mass $M$ containing an average density of $\bar{\rho}_{\rm M}(z)$. The latter can be computed using the $\Lambda$CDM linear power spectrum (for both GR and $f(R)$ gravity) with the publicly available code \textsc{camb} \citep{Lewis:1999bs}. 

The $f(R)$ effects are incorporated through $\delta_{\rm c}$: in GR, this is given by:
\begin{equation}
    \delta_{\rm c}^{\rm GR}(z)\approx\frac{3}{20}(12\pi)^{\frac{2}{3}}\left[1 + 0.0123\log_{10}\Omega_{\rm M}(z)\right],
    \label{eq:delta_c_gr}
\end{equation}
while in $f(R)$ gravity it can be expressed as:
\begin{equation}
    \delta_{\rm c}^{\rm eff}(M, z)\equiv\epsilon(M, z)\times\delta_{\rm c}^{f(R)}(M, z).
    \label{eq:delta_eff}
\end{equation}
The function $\delta_{\rm c}^{f(R)}(M,z)$ is the prediction of the linearly extrapolated threshold density for spherical collapse in $f(R)$ gravity. This treats haloes and their surrounding environment as co-centred spherically symmetric top-hat overdensities (note the environment can be underdensities) which are co-evolved from an initial time to the time of collapse. This procedure, which is based on the method developed by \citet{LE2012,Lombriser:2013wta}, takes into account both the mass-dependent self-screening and the environmental screening of the fifth force. However, while giving qualitatively correct predictions, the method is unable to very accurately capture the complex nonlinear dynamics of structure formation in $f(R)$ gravity. This limitation is accounted for using the correction factor $\epsilon(M,z)$, which \citet{Cataneo:2016iav} modelled and fitted using dark-matter-only simulations. Their best-fit model can accurately reproduce the $f(R)$ enhancement of the HMF for redshifts $0.0\leq z\leq0.5$ and field strengths $10^{-6}\leq |f_{R0}|\leq10^{-4}$.

\begin{figure}
\centering
\includegraphics[width=\columnwidth]{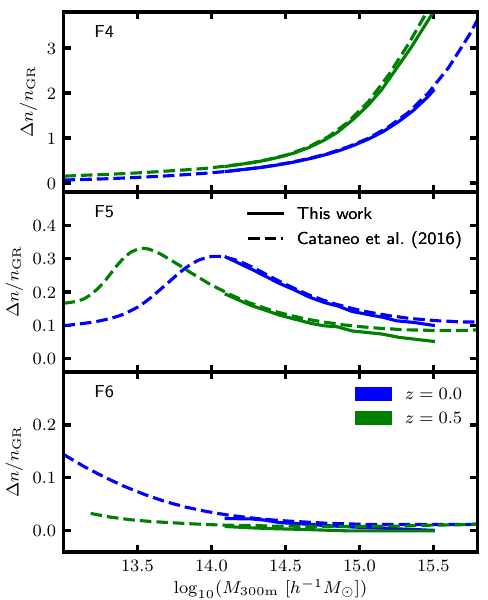}
\caption{[{\it Colour Online}] Halo mass function enhancement, $\Delta n/n_{\rm GR} = n_{f(R)}/n_{\rm GR}-1$, in $f(R)$ gravity with respect to GR as a function of the halo mass. The solid lines show the predictions from our 5D interpolation of $\delta_{\rm c}^{f(R)}$ (see Sec.~\ref{sec:background:hmf}) and the dashed lines show the results from \citet{Cataneo:2016iav}. The predictions have been generated using the WMAP9 cosmological parameters and $f(R)$ models F4 (\textit{top row}), F5 (\textit{middle row}) and F6 (\textit{bottom row}), at redshifts 0 (\textit{blue lines} and 0.5 (\textit{green lines}).}
\label{fig:hmf_enhancement}
\end{figure}

For this work, we have evaluated $\delta_{\rm c}^{f(R)}$ on a grid of $M$, $z$, $\Omega_{\rm M}$, $\sigma_8$ and $f_{R0}$, and obtained the relation $\delta_{\rm c}^{f(R)}(M,z,\Omega_{\rm M},\sigma_8,f_{R0})$ using 5D interpolation. For a given set of cosmological and $f(R)$ parameters, we can use this to predict $\delta_{\rm c}^{f(R)}(M,z)$, which can then be used to predict $\delta_{\rm c}^{\rm eff}(M,z)$ using the model for $\epsilon(M,z)$ taken from \citet{Cataneo:2016iav}. The $f(R)$ enhancement of the HMF is given by the ratio between $n_{\rm ST}|_{f(R)}$ and $n_{\rm ST}|_{\rm GR}$, which are evaluated using $\delta_{\rm c}=\delta_{\rm c}^{\rm eff}$ and $\delta_{\rm c}=\delta_{\rm c}^{\rm GR}$, respectively.

For illustrative purposes, we show, in Fig.~\ref{fig:hmf_enhancement}, our predictions of the HMF enhancement as a function of the halo mass for F6, F5 and F4 at redshifts 0.0 and 0.5. We also show the predictions from \citet{Cataneo:2016iav} as a comparison. Both sets of predictions assume the 9-year WMAP cosmological parameter estimates \citep{2013ApJS..208...19H}. There are some small differences between the two sets of predictions, which are likely caused by subtle differences in the calculations of $\delta_{\rm c}^{f(R)}$. The largest difference is observed at $M_{\rm 300m}\gtrsim10^{15}h^{-1}M_{\odot}$ for F5 at $z=0.5$. We note that the enhancement is expected to drop to zero at high masses where haloes become completely screened, therefore the behaviour of the solid lines here 
appears to be physically reasonable. We also note that we set the enhancement to zero wherever our calculations predict a negative (unphysical) enhancement. This is the case for $M_{\rm 300m}\gtrsim10^{15}h^{-1}M_{\odot}$ for F6 at $z=0.5$.

\section{Methods}
\label{sec:methods}

In this section, we describe the main components of our constraint pipeline, including the mass function predictions (Sec.~\ref{sec:methods:hmf}), the observable-mass scaling relation (Sec.~\ref{sec:methods:scaling_relation}), the mock cluster catalogues (Sec.~\ref{sec:methods:mock}) and the MCMC sampling (Sec.~\ref{sec:methods:likelihood}).

\subsection{Theoretical mass function}
\label{sec:methods:hmf}

In order to make constraints using cluster number counts, it is necessary to have a parameter-dependent theoretical model for the HMF. For this work, we start with a GR HMF and apply the $f(R)$ enhancement using: 
\begin{equation}
    n^{f(R)} = n^{\rm GR}\times\frac{n_{\rm ST}|_{f(R)}}{n_{\rm ST}|_{\rm GR}},
    \label{eq:fr_hmf}
\end{equation}
where the ratio is computed using the \citet{Sheth:1999mn} prescription, as described in Sec.~\ref{sec:background:hmf}, and we have chosen the \citet{Tinker:2008ff} calibration for $n^{\rm GR}$. 

Before Eq.~(\ref{eq:fr_hmf}) can be applied, the halo mass definition must be considered. As mentioned in Sec.~\ref{sec:background:hmf}, the model for the ratio in Eq.~(\ref{eq:fr_hmf}) was calibrated by \citet{Cataneo:2016iav} using overdensity $\Delta=300\Omega_{\rm M}(z)$; however, with the framework in Fig.~\ref{fig:flow_chart}, we hope to use data from SZ and X-ray surveys, which often measure cluster properties with overdensity $500$. Therefore, it is necessary to convert the HMF between these two definitions. 

\begin{figure}
\centering
\includegraphics[width=\columnwidth]{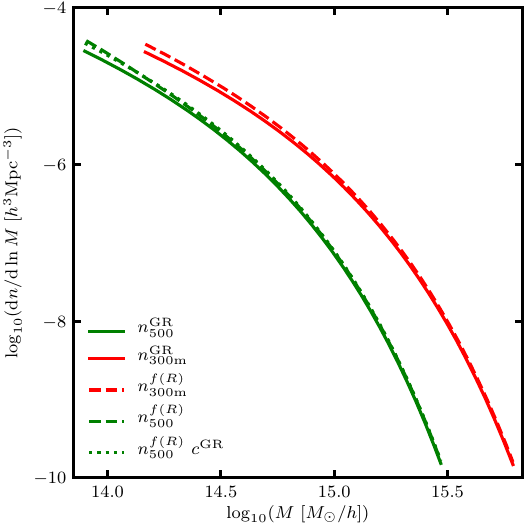}
\caption{[{\it Colour Online}] Halo mass function in GR (\textit{solid lines}) and F5 (\textit{dashed lines}), with the mass defined using spherical overdensities $500$ (\textit{green lines}) and $300\Omega_{\rm M}(z)$ (\textit{dashed lines}). The mass conversions and $f(R)$ enhancement have been applied as described in Sec.~\ref{sec:methods:hmf}; the green dotted line shows the F5 HMF prediction that results from neglecting the $f(R)$ enhancement of the halo concentration in the mass conversion $300\Omega_{\rm M}(z)\rightarrow500$.}
\label{fig:hmf_conversion}
\end{figure}

In Fig.~\ref{fig:hmf_conversion}, we show each step of the mass conversion procedure for the F5 model at $z=0$. We start with the \citet{Tinker:2008ff} HMF with overdensity $500$ ($n_{500}^{\rm GR}$), which we compute using the python package \textsc{hmf} \citep{Murray:2013qza}, and convert this to overdensity $300\Omega_{\rm M}(z)$ ($n_{\rm 300m}^{\rm GR}$) using the \citet{Duffy:2008pz} concentration-mass-redshift relation. We then apply the $f(R)$ enhancement using Eq.~(\ref{eq:fr_hmf}) to get $n_{\rm 300m}^{f(R)}$. Finally, to convert this back to overdensity $500$ ($n_{500}^{f(R)}$), we use the $f(R)$ concentration-mass-redshift relation, which is computed by applying the concentration enhancement, given by Eq.~(\ref{eq:c_model}), to the \citet{Duffy:2008pz} relation. We also show, with the dotted green line, the prediction with the concentration enhancement neglected; the effect here is quite small, since cluster-size haloes are mostly screened in F5. For further details of the formulae used to convert the halo mass and the HMF from one mass definition to another, we refer the reader to Appendix \ref{sec:appendix:mass_conversions}.

The final result $n_{500}^{f(R)}(M_{500})$ provides the theoretical prediction of the cluster abundance in $f(R)$ gravity. This is computed following the above steps for each set of parameter values sampled by our MCMC pipeline. We note that our mass conversions are evaluated assuming an NFW profile, which has also been used in previous cluster tests of $f(R)$ gravity \citep[e.g.,][]{PhysRevD.92.044009}. However, this may not provide an accurate description for haloes that are not dynamically relaxed and it does not account for the effects of baryons on the total mass profile. We plan to investigate the latter effect using clusters identified from the realistic full hydrodynamical simulations in $f(R)$ gravity described in \citet{Mitchell:2021ter}. However, we remark here that the main use of the concentration-mass relation in our pipeline is to perform mass conversions as described above, and so it would not be strictly needed if a theoretical HMF for the required mass definition $M_\Delta$ ($M_{500}$ for this paper) is already in place.

\subsection{Observable-mass scaling relation}
\label{sec:methods:scaling_relation}

As discussed in Sec.~\ref{sec:background:scaling_relations}, the $f(R)$ scaling relation can be computed by simply rescaling a GR relation using our model for the dynamical mass enhancement. For the GR relation, we adopt the power-law mapping between $Y_{\rm SZ}$ and the halo mass calibrated by the Planck Collaboration \citep{Planck_SZ_cluster}:
\begin{equation}
    E^{-\beta}(z)\left[\frac{D_{\rm A}^2(z)\bar{Y}_{500}}{10^{-4}{\rm Mpc}^2}\right] = Y_{\star}\left[\frac{h}{0.7}\right]^{-2+\alpha}\left[\frac{(1-b)M_{500}}{6\times10^{14}M_{\odot}}\right]^{\alpha},
    \label{eq:planck_ysz}
\end{equation}
where $E(z)=H(z)/H_0$ and $D_{\rm A}(z)$ is the angular diameter distance. This includes parameters $\beta$ for the $z$-evolution, $Y_{\star}$ for the normalisation and $\alpha$ for the power-law slope with respect to the mass. It also includes a bias parameter $(1-b)$ which accounts for differences between the X-ray determined masses used in the calibration, which are subject to hydrostatic equilibrium bias, and the true mass. Planck have also provided the following formula for the intrinsic lognormal scatter of the relation:
\begin{equation}
    P(\log Y_{500}) = \frac{1}{\sqrt{2\pi}\sigma_{\log Y}}\exp\left[-\frac{\log^2(Y_{500}/\bar{Y}_{500})}{2\sigma_{\log Y}^2}\right],
    \label{eq:intrinsic_scatter}
\end{equation}
where $\sigma_{\log Y}$ is a fixed spread.

We assume a fixed value of 0.8 for the hydrostatic equilibrium bias parameter, which is consistent with the range 0.7 to 1.0 adopted by Planck, and we treat $Y\equiv D_{\rm A}^2(z)Y_{\rm SZ}$ as the cluster SZ observable, rather than $Y_{\rm SZ}$. This leaves four scaling relation parameters which are allowed to vary in our MCMC sampling. We adopt the following Gaussian priors from Planck: $\log Y_{\star}=-0.19\pm0.02$, $\alpha=1.79\pm0.08$, $\beta=0.66\pm0.50$ and $\sigma_{\log Y}=0.075\pm0.010$. 

To obtain the $f(R)$ scaling relation $Y_{\rm SZ}^{f(R)}(M_{500})$ from the above $Y_{\rm SZ}^{\rm GR}(M_{500})$ relation, we rescale the right-hand side of Eq.~(\ref{eq:planck_ysz}) by the mass ratio $\mathcal{R}$, which is predicted using Eq.~(\ref{eq:mdyn_enhancement}) with scatter given by Eq.~(\ref{eq:scatter_model}). This rescaling is based on Eq.~(\ref{eq:ysz_mapping_true}), which means that the mass $M_{500}$ in the expressions $Y_{\rm SZ}^{f(R)}(M_{500})$ and $Y_{\rm SZ}^{\rm GR}(M_{500})$ above is the true mass; we note that, although the Planck masses were originally determined using X-ray measurements, the value $(1-b)=0.8$ assumed for the mass bias is consistent with weak lensing measurements \citep[e.g.,][]{Hoekstra:2015gda}. 

Finally, we note that the scaling relation adopted in this work is intended to be representative of general scaling relations between the mass and SZ and X-ray observables, not just the Planck $Y_{\rm SZ}(M_{500})$ relation. This justifies our decision to encapsulate $D_{\rm A}^2(z)$ in the cluster observable and to fix the hydrostatic equilibrium bias; indeed, scaling relations for other observables -- for example, the SZ significance and the $Y_{\rm X}$ parameter -- do not include the function $D_{\rm A}^2(z)$ or a bias parameter \citep[e.g.,][]{SPT:2016izt,SPT:2018njh}. Regardless of the observable, the main purpose of this work is to check that our constraint pipeline can give reasonable constraints of $f_{R0}$ using a realistic scaling relation which includes both intrinsic scatter and the $f(R)$ enhancement. It would be very straightforward to adapt this pipeline for other cluster observables, or for more than one cluster observable.

\subsection{Mock catalogues}
\label{sec:methods:mock}

We test our framework (Fig.~\ref{fig:flow_chart}) using mock cluster catalogues in place of observational data. We have generated mocks for both the GR and F5 models, using fiducial cosmological parameter values based on the Planck 2018 CMB constraints \citep{Planck:2018vyg}: $(\Omega_{\rm M},\sigma_8,h,\Omega_{\rm b},n_{\rm s})=(0.3153,0.8111,0.6736,0.04931,0.9649)$. For the scaling relation parameters, we assume the central values of the Gaussian priors listed in Sec.~\ref{sec:methods:scaling_relation}. 

To generate the mocks, we first compute the predicted count per unit mass per unit redshift:
\begin{equation}
    \frac{{\rm d}N}{{\rm d}z{\rm d}\ln M} = \frac{{\rm d}n}{{\rm d}\ln M}\times\frac{{\rm d}V_{\rm c}(z)}{{\rm d}z},
    \label{eq:count_density}
\end{equation}
where $V_{\rm c}(z)$ is the comoving volume enclosed by the survey area between redshifts 0 and $z$ and the first term is the theoretical HMF $n_{500}^{f(R)}$, which is computed as described in Sec.~\ref{sec:methods:hmf} for the fiducial cosmology. For this work, we assume a survey area of 5000 deg$^2$ and a maximum redshift of $z=0.5$, which is the upper redshift used to calibrate the $f(R)$ enhancement of the HMF (Sec.~\ref{sec:background:hmf}). In the future, we plan to develop models of the $f(R)$ HMF that work for a wider redshift range, which will be applicable to real cluster survey data.

The predicted number of clusters is:
\begin{equation}
    N_{\rm tot} = \int_{0.0}^{0.5}{\rm d}z\int_{-\infty}^{\infty}{\rm d}\ln M\frac{{\rm d}N}{{\rm d}z{\rm d}\ln M}.
\end{equation}
For each mock, we randomly draw the masses and redshifts of $N_{\rm tot}$ clusters using ${\rm d}N/{\rm d}z{\rm d}\ln M$, which is effectively a probability density. For each cluster $i$, we then draw a mass ratio $\mathcal{R}_i$ using a normal distribution with mean given by Eq.~(\ref{eq:mdyn_enhancement}) and standard deviation given by Eq.~(\ref{eq:scatter_model}). The intrinsic observable $Y'_i(=D_{\rm A}^2Y_{500,i})$ of each cluster is then drawn using the lognormal distribution given by Eq.~(\ref{eq:intrinsic_scatter}), where $\bar{Y}_{500}$ is computed using Eq.~(\ref{eq:planck_ysz}) and rescaled by $\mathcal{R}_i$. 

We assume a fixed $1\sigma$ measurement uncertainty of $10\%$. The measured observable $Y_i$ is therefore drawn from a normal distribution with mean $Y'_i$ and standard deviation $0.1Y'_i$. We note that this choice of a fixed fractional uncertainty is intended to keep our calculations simple and general (for example, a more complicated model may be specific to a particular  observational survey). We have also considered $5\%$ and $20\%$ uncertainties and have found that the inferred parameter constraints do not significantly differ, suggesting that this uncertainty is not the dominant source of error in the constraint pipeline (e.g., compared to the intrinsic scatters in the cluster scaling relation or the $f(R)$ dynamical mass enhancement).

Finally, we remove all clusters for which $Y_i$ is below some observational flux limit $Y_{\rm cut}$. For the main results of this work, we use $Y_{\rm cut}=1.5\times10^{-5}{\rm Mpc}^2$; however, we will also discuss the effects of using cuts $10^{-5}{\rm Mpc}^2$, $2\times10^{-5}{\rm Mpc}^2$ and $2.5\times10^{-5}{\rm Mpc}^2$. For each mock, we store only the cluster redshift $z_i$ (which is assumed to have no error) and the measured observable $Y_i$.

\begin{figure}
\centering
\includegraphics[width=\columnwidth]{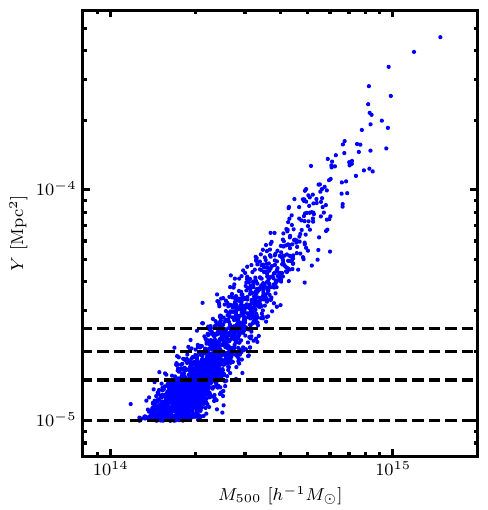}
\caption{[{\it Colour Online}] SZ $Y$-parameter as a function of the halo mass for clusters from a GR mock catalogue with observational flux limit $Y_{\rm cut}=10^{-5}{\rm Mpc}^2$. The dashed lines indicate the cuts $10^{-5}{\rm Mpc}^2$, $1.5\times10^{-5}{\rm Mpc}^2$, $2\times10^{-5}{\rm Mpc}^2$ and $2.5\times10^{-5}{\rm Mpc}^2$.}
\label{fig:mock_sr}
\end{figure}

An F5 mock with $Y_{\rm cut}=1.5\times10^{-5}{\rm Mpc}^2$ contains $\sim1350$ clusters. Generating a GR mock is more straightforward, since there is no need to include the $f(R)$ enhancements of the HMF or the scaling relation. In this case, there are $\sim1150$ clusters for $Y_{\rm cut}=1.5\times10^{-5}{\rm Mpc}^2$. For illustrative purposes, in Fig.~\ref{fig:mock_sr} we show the measured $Y$-parameters of the clusters as a function of the mass $M_{500}$ for a GR mock with $Y_{\rm cut}=10^{-5}{\rm Mpc}^2$. Horizontal dashed lines are included to indicate the four flux thresholds considered in this work, to give an idea of the mass range of clusters found above each.

\subsection{MCMC sampling}
\label{sec:methods:likelihood}

For our parameter constraints, we use the unbinned Poisson likelihood \citep[e.g.,][]{Artis:2021tjj}:
\begin{equation}
    \ln\mathcal{L} = -\int {\rm d}z{\rm d}Y\frac{{\rm d}N}{{\rm d}z{\rm d}Y}(z,Y) + \sum_i\ln\frac{{\rm d}N}{{\rm d}z{\rm d}Y}(z_i,Y_i),
    \label{eq:log_likelihood}
\end{equation}
where the first term represents the predicted cluster count and the second term is a summation performed over all mock clusters. The expression ${\rm d}N/{\rm d}z{\rm d}Y$ represents the theoretical prediction of the count per unit $z$ per unit $Y$. 

Since our theoretical HMF is defined in terms of the mass $M$ rather than the observable $Y$, it is more convenient to re-express the first term with an integral over $\ln M$ \citep[e.g.,][]{SPT:2016izt}:
\begin{equation}
    \begin{split}
    &-\int_{0.0}^{0.5}{\rm d}z\int_{Y_{\rm cut}}^{\infty}{\rm d}Y\frac{{\rm d}N}{{\rm d}z{\rm d}Y}(z,Y)\\
    &= -\int_{0.0}^{0.5}{\rm d}z\int_{-\infty}^{\infty}{\rm d}\ln MP(Y>Y_{\rm cut}|M,z)\frac{{\rm d}N}{{\rm d}z{\rm d}\ln M}(M,z),
    \end{split}
    \label{eq:predicted_count}
\end{equation}
where ${\rm d}N/{\rm d}z{\rm d}\ln M$ can be computed using the Eq.~(\ref{eq:count_density}), and the redshift integral is evaluated between $z=0$ and the maximum redshift $z=0.5$ of the mock. $P(Y>Y_{\rm cut}|M,z)$ represents the probability that, for a given mass and redshift, the measured $Y$-parameter exceeds the flux threshold. This depends on both the measurement uncertainty and the intrinsic log-normal scatter of $Y$:
\begin{equation}
    P(Y > Y_{\rm cut}|M,z) = \int_{-\infty}^{\infty}{\rm d}\ln Y' P(Y > Y_{\rm cut}|Y')P(Y'|M,z),
    \label{eq:cut_prob}
\end{equation}
where $P(Y > Y_{\rm cut}|Y')$ is the probability that the measured value $Y$ exceeds $Y_{\rm cut}$, given an intrinsic value $Y'$, and $P(Y'|M,z)$ is the probability density of a cluster having intrinsic value $Y'$ given that it has mass $M$ and redshift $z$. As discussed in Sec.~\ref{sec:methods:mock}, the mocks use a fixed measurement uncertainty of $10\%$, which means that the former can be estimated using a normal distribution with mean $Y'$ and standard deviation $0.1Y'$. The probability density $P(Y'|M,z)$ is more complicated, since this depends both on the intrinsic scatter of the $Y(M)$ scaling relation and the scatter of the mass ratio $\mathcal{R}$:
\begin{equation}
    P(Y'|M,z) = \int_{1}^{4/3}{\rm d}\mathcal{R} P(Y'|\bar{Y}(M,z,\mathcal{R}))P(\mathcal{R}|M,z),
    \label{eq:ratio_integral}
\end{equation}
where $P(\mathcal{R}|M,z)$ is the probability density of a cluster having mass ratio $\mathcal{R}$ given that it has mass $M$ and redshift $z$. This is computed using a normal distribution with mean given by Eq.~(\ref{eq:mdyn_enhancement}) and standard deviation given by Eq.~(\ref{eq:scatter_model}). The other probability density, $P(Y'|\bar{Y}(M,z,\mathcal{R}))$, is computed using Eq.~(\ref{eq:intrinsic_scatter}), with $\bar{Y}$ calculated using Eq.~(\ref{eq:planck_ysz}) and rescaled by a factor of $\mathcal{R}$. Together, Eqs.~(\ref{eq:predicted_count})-(\ref{eq:ratio_integral}) form a 4D integral, which we compute using a fixed grid in ($\ln M,z,\ln Y,\mathcal{R}$).

For the second term in Eq.~(\ref{eq:log_likelihood}), we can again re-express into a form that depends on ${\rm d}N/({\rm d}z{\rm d}\ln M)$ using:
\begin{equation}
    \begin{split}
    \frac{{\rm d}N}{{\rm d}z{\rm d}Y}(z_i,Y_i) = &\int{\rm d}\ln Y'\int{\rm d}\ln M'\\ 
    &\times P(Y_i|Y')P(Y'|M',z_i)\frac{{\rm d}N}{{\rm d}z{\rm d}\ln M'}(M',z_i),
    \end{split}
    \label{eq:second_term}
\end{equation}
where the probability density functions $P(Y_j|Y')$ and $P(Y'|M',z_i)$ represent the measurement uncertainty and intrinsic scatter, respectively. The latter is computed using Eq.~(\ref{eq:ratio_integral}), meaning that Eq.~(\ref{eq:second_term}) is really a 3D integral. We compute this for each mock cluster using a fixed grid in ($\ln Y',\ln M',\mathcal{R}$), then evaluate the sum in Eq.~(\ref{eq:log_likelihood}).

We have used the python package \textsc{emcee} \citep{2013PASP..125..306F} for the MCMC sampling. For all of the results discussed in this work, we have used 28 walkers each travelling 2700 steps (we discard the first 600 steps to ensure that the chains are well converged). At each step, the log-likelihood is computed for the sampled parameters as described above. In addition to the $f_{R0}$ parameter, the cosmological parameters $\Omega_{\rm M}$ and $\sigma_8$ and the four scaling relation parameters $Y_{\star}$, $\alpha$, $\beta$ and $\sigma_{\log Y}$ are sampled. For the cosmological parameters, we adopt uniform (flat) priors $\log_{10}|f_{R0}|\in[-7,-4]$ and $\sigma_8\in[0.60,0.95]$, and for $\Omega_{\rm M}$ we use either a flat prior $\Omega_{\rm M}\in[0.15,0.50]$ or a Gaussian prior $\Omega_{\rm M}=0.3153\pm0.0073$ which is based on the Planck 2018 CMB constraints \citep{Planck:2018vyg}. For the scaling relation parameters, we adopt the Gaussian priors listed in Sec.~\ref{sec:methods:scaling_relation}.

The flat prior $[-7,-4]$ for $\log_{10}|f_{R0}|$ extends beyond the range $[-6,-4]$ used to calibrate the HMF enhancement model \citep{Cataneo:2016iav}. For sampled values in the range $-7\leq\log_{10}|f_{R0}|\leq-6$, we first calculate the HMF enhancement for $\log_{10}|f_{R0}|=-6$, then linearly interpolate between $|f_{R0}|=0$ (GR) and $|f_{R0}|=10^{-6}$ to estimate the enhancement. For example, this means that the estimated enhancement for $|f_{R0}|=10^{-7}$ would be $10\%$ of the enhancement for $|f_{R0}|=10^{-6}$. We note that, because clusters are expected to be completely screened for this range of $\log_{10}|f_{R0}|$ values, it is not necessary to use a physically accurate method here, so long as the predicted enhancement lies between GR and F6. We use a similar approach to estimate the dynamical mass enhancement for this range of $\log_{10}|f_{R0}|$, where, again, the enhancement is very close to zero anyway.

\section{Results}
\label{sec:results}

In this section, we discuss the main results of this work. In Sec.~\ref{sec:results:gr_pipeline}, we use a GR mock to check that our pipeline can give reasonable constraints of the $\Lambda$CDM and scaling relation parameters. Then, in Sec.~\ref{sec:results:fr_pipeline}, we use our full pipeline to constrain the $f_{R0}$ parameter of $f(R)$ gravity, using a combination of GR and F5 mocks.

\subsection{GR pipeline}
\label{sec:results:gr_pipeline}

\begin{figure*}
\centering
\includegraphics[width=0.9\textwidth]{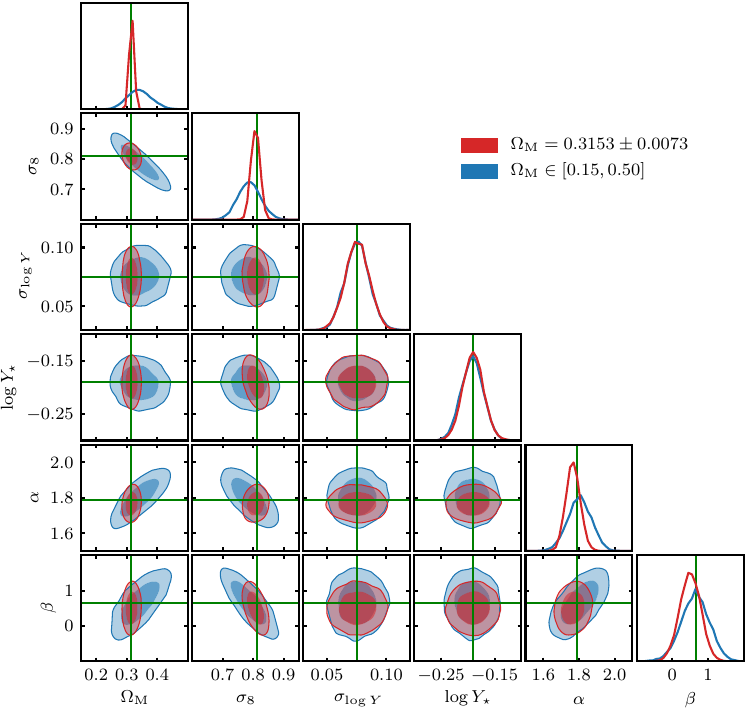}
\caption{[{\it Colour Online}] Parameter constraints using our GR pipeline, which does not include $f(R)$ enhancements of the HMF and the scaling relation (see Sec.~\ref{sec:methods}), using a GR mock with observational flux threshold $Y_{\rm cut}=1.5\times10^{-5}{\rm Mpc}^2$. The two sets of constraints are generated using a flat prior $[0.15,0.50]$ (\textit{blue}) and a Gaussian prior $0.3153\pm0.0073$ (\textit{red}) in $\Omega_{\rm M}$. The dark and light regions of the contours represent 68\% and 95\% confidences, respectively. The distributions of the sampled parameter values are shown in the top panels of each column, with the mean and standard deviation of each parameter quoted in Table~\ref{table:gr_pipeline}. The fiducial cosmological parameter values of the GR mock are indicated by the green lines.}
\label{fig:gr_pipeline}
\end{figure*}

\begin{table*}
\centering

\small
\begin{tabular}{ c@{\hskip 0.5in}cc@{\hskip 0.5in}cc@{\hskip 0.5in}cc } 
 \toprule
 
  & & & \multicolumn{2}{c}{Flat $\Omega_{\rm M}$ prior} & \multicolumn{2}{c}{Gaussian $\Omega_{\rm M}$ prior} \\
 Parameter & Fiducial value & Prior & 68\% range & $\mathcal{L}_{\rm max}$ & 68\% range & $\mathcal{L}_{\rm max}$ \\

 \midrule

 $\Omega_{\rm M}$ & $0.3153$ & --- & $0.34\pm0.04$ & $0.3384$ & $0.316\pm0.007$ & $0.3175$ \\ 
 $\sigma_8$ & $0.8111$ & $[0.60,0.95]$ & $0.79\pm0.04$ & $0.7888$ & $0.808\pm0.015$ & $0.8037$ \\
 $\sigma_{\log Y}$ & $0.075$ & $0.075\pm0.010$ & $0.076\pm0.010$ & $0.076$ & $0.076\pm0.010$ & $0.073$ \\
 $\log Y_{\star}$ & $-0.19$ & $-0.19\pm0.02$ & $-0.19\pm0.02$ & $-0.19$ & $-0.19\pm0.02$ & $-0.19$\\
 $\alpha$ & $1.79$ & $1.79\pm0.08$ & $1.80\pm0.07$ & $1.80$ & $1.77\pm0.04$ & $1.76$\\
 $\beta$ & $0.66$ & $0.66\pm0.50$ & $0.6\pm0.4$ & $0.63$ & $0.5\pm0.3$ & $0.58$\\
 
 \bottomrule
 
\end{tabular}

\caption{Parameter constraints using our GR pipeline. The mean and standard deviation are quoted (68\% range) along with the parameter combinations giving the highest log-likelihood ($\mathcal{L}_{\rm max}$). The constraints correspond to the distributions shown in Fig.~\ref{fig:gr_pipeline}.}
\label{table:gr_pipeline}

\end{table*}

In order to verify that our pipeline can give reasonable $\Lambda$CDM constraints and successfully account for the intrinsic scatter of the $Y(M_{500})$ relation and measurement uncertainty in the mock, we first test our `GR pipeline', where the $f(R)$ corrections to the HMF and scaling relation are excluded. We show the constraints, which have been inferred using a GR mock with $Y_{\rm cut}=1.5\times10^{-5}{\rm Mpc}^2$, in Fig.~\ref{fig:gr_pipeline}. The blue contours are obtained using the flat prior $\Omega_{\rm M}\in[0.15,0.50]$, while the red contours are obtained using the Gaussian prior $\Omega_{\rm M}=0.3153\pm0.0073$ from Planck 2018. 

For the flat $\Omega_{\rm M}$ prior, the contours are in good agreement with the fiducial parameter values, which are indicated by the green lines. In the top panel of each column, we show the marginalised distributions of each parameter, with the mean and standard deviation quoted in Table~\ref{table:gr_pipeline}. In Table~\ref{table:gr_pipeline}, we also show the combination of parameters that gave the highest log-likelihood during the sampling ($\mathcal{L}_{\rm max}$); these can be thought of as the `most likely' set of values. The distributions of the scaling relation parameters closely match the Gaussian priors. Meanwhile, the constraints $0.34\pm0.04$ for $\Omega_{\rm M}$ and $0.79\pm0.04$ for $\sigma_8$ -- while still within $1\sigma$ agreement -- are slightly offset from the fiducial values, and the same goes for the highest-likelihood values 0.34 and 0.79. As shown by the constraints in red, using a tighter Gaussian prior in $\Omega_{\rm M}$ results in narrower contours and constraints $\Omega_{\rm M}=0.316\pm0.007$ and $\sigma_8=0.808\pm0.015$ which match the fiducial values more closely.

The initial offset of the $\Omega_{\rm M}$ and $\sigma_8$ constraints from the fiducial values is caused by a well-known degeneracy between these two parameters: increasing either of these will boost the predicted amplitude of the HMF. Therefore, the effects of increasing (decreasing) $\Omega_{\rm M}$ and decreasing (increasing) $\sigma_8$ on the HMF can roughly cancel out. This causes the elongated shape of the blue $\Omega_{\rm M}$-$\sigma_8$ contour. 

We also observe degeneracies between $\Omega_{\rm M}$, $\sigma_8$, $\alpha$ and $\beta$. One explanation for this is that $\alpha$ and $\beta$ can also affect the predicted HMF. For example, increasing $\alpha$ (i.e., increasing the slope of the $Y(M)$ scaling relation) will cause the predicted $Y$-parameter to be reduced for clusters with $0.8M_{500}<6\times10^{14}M_{\odot}$ (since $(1-b)^{-1}6\times10^{14}M_\odot$ is the pivot mass of the power-law function in Eq.~\eqref{eq:planck_ysz}), which includes the majority of clusters in our mocks (see Fig.~\ref{fig:mock_sr}). This means that fewer clusters will be predicted to have $Y>Y_{\rm cut}$, and therefore the inferred cluster count will be lower, which can be countered by a larger $\Omega_{\rm M}$. The effects of changing $\alpha$, $\beta$, $\Omega_{\rm M}$ and $\sigma_8$ may balance out overall, giving rise to the observed degeneracies in the blue contours of Fig.~\ref{fig:gr_pipeline}. 

By adopting the tighter $\Omega_{\rm M}$ prior, these degeneracies appear to be mostly eliminated. This shows the importance of accurate independent measurements of $\Omega_{\rm M}$ in the use of galaxy cluster number counts to constrain cosmological models and parameters.

\subsection{\boldmath \texorpdfstring{$f(R)$}{f(R)} pipeline}
\label{sec:results:fr_pipeline}

\begin{figure*}
\centering
\includegraphics[width=\textwidth]{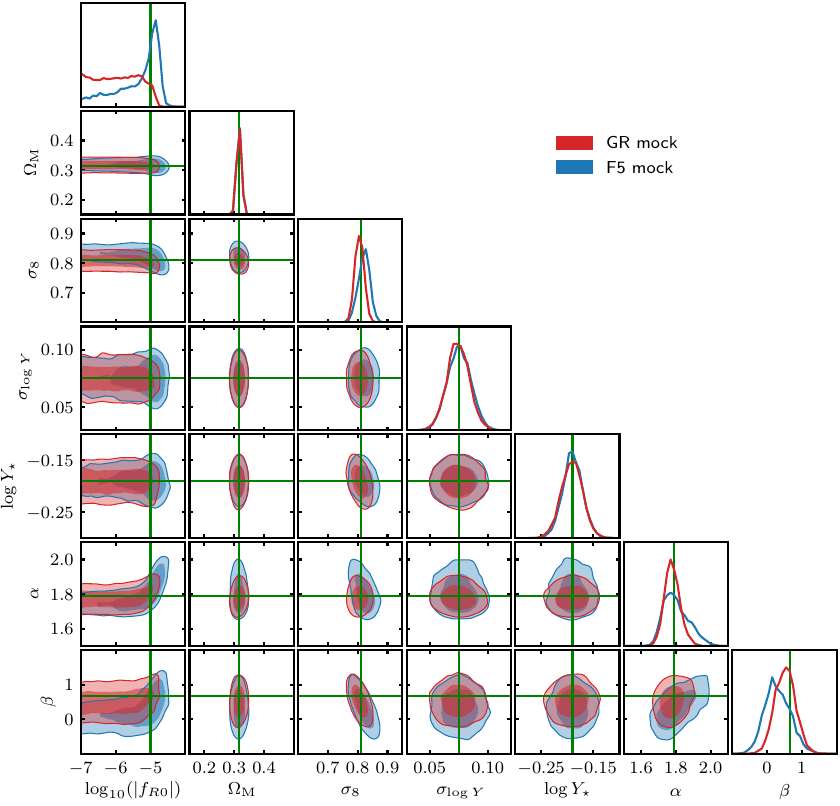}
\caption{[{\it Colour Online}] Parameter constraints using our full $f(R)$ pipeline, as detailed in Sec.~\ref{sec:methods}, using GR (\textit{red}) and F5 (\textit{blue}) mocks with observational flux threshold $Y_{\rm cut}=1.5\times10^{-5}{\rm Mpc}^2$. The dark and light regions of the contours represent 68\% and 95\% confidences, respectively. The distributions of the sampled parameter values are shown in the top panels of each column, with the mean and standard deviation quoted in Table~\ref{table:fr_pipeline}. The fiducial cosmological parameter values of the mocks are indicated by the green lines, including the value $\log_{10}|f_{R0}|=-5$ for the F5 mock.}
\label{fig:full_fr_pipeline}
\end{figure*}

\begin{table*}
\centering

\small
\begin{tabular}{ c@{\hskip 0.5in}cc@{\hskip 0.5in}cc@{\hskip 0.5in}cc } 
 \toprule
 
  & & & \multicolumn{2}{c}{GR constraints} & \multicolumn{2}{c}{F5 constraints} \\
 Parameter & Fiducial value & Prior & 68\% range & $\mathcal{L}_{\rm max}$ & 68\% range & $\mathcal{L}_{\rm max}$ \\

 \midrule

 $\log_{10}|f_{R0}|$ & --- & $[-7,-4]$ & $\leq-5.56$ & $-6.75$ & $-5.1^{+0.3}_{-1.0}$ & $-4.92$ \\ 
 $\Omega_{\rm M}$ & $0.3153$ & $0.3153\pm0.0073$ & $0.316\pm0.007$ & $0.317$ & $0.316\pm0.008$ & $0.313$ \\ 
 $\sigma_8$ & $0.8111$ & $[0.60,0.95]$ & $0.806\pm0.015$ & $0.806$ & $0.821\pm0.019$ & $0.815$ \\
 $\sigma_{\log Y}$ & $0.075$ & $0.075\pm0.010$ & $0.075\pm0.010$ & $0.072$ & $0.075\pm0.010$ & $0.079$ \\
 $\log Y_{\star}$ & $-0.19$ & $-0.19\pm0.02$ & $-0.19\pm0.02$ & $-0.19$ & $-0.190\pm0.019$ & $-0.18$\\
 $\alpha$ & $1.79$ & $1.79\pm0.08$ & $1.78\pm0.04$ & $1.77$ & $1.80\pm0.07$ & $1.82$\\
 $\beta$ & $0.66$ & $0.66\pm0.50$ & $0.5\pm0.3$ & $0.51$ & $0.3\pm0.4$ & $0.44$\\
 
 \bottomrule
 
\end{tabular}

\caption{Parameter constraints using our full $f(R)$ pipeline. The 68\% range columns show the mean and standard deviation for all parameters other than $\log_{10}|f_{R0}|$; for the latter, the 68\% upper bound is shown for the GR mock constraints and the median and 68-percentile is shown the F5 mock constraints. The parameter combinations giving the highest log-likelihood ($\mathcal{L}_{\rm max}$) are also shown. The constraints correspond to the distributions shown in Fig.~\ref{fig:full_fr_pipeline}.}
\label{table:fr_pipeline}

\end{table*}

We now test the full $f(R)$ gravity constraint pipeline, which includes the $f(R)$ effects on the HMF and scaling relation, as described in Secs.~\ref{sec:methods:hmf} and \ref{sec:methods:scaling_relation}. In Fig.~\ref{fig:full_fr_pipeline}, we show the constraints inferred using GR and F5 mocks with $Y_{\rm cut}=1.5\times10^{-5}{\rm Mpc}^2$. For these results, we use the Gaussian prior of $\Omega_{\rm M}$ in order to prevent the $\Omega_{\rm M}$--$\sigma_8$ degeneracy observed in Fig.~\ref{fig:gr_pipeline}. As we will show in Sec.~\ref{sec:bias}, using a flat prior for $\Omega_{\rm M}$ can otherwise lead to biased constraints of $\log_{10}|f_{R0}|$.

For the constraints obtained from the GR mock, which are indicated by the red contours in Fig.~\ref{fig:full_fr_pipeline}, the $\log_{10}|f_{R0}|$ posterior distribution is roughly uniform for the range $-7\leq\log_{10}|f_{R0}|\lesssim-5$ and drops to zero for $\log_{10}|f_{R0}|\gtrsim-5$. This rules out $f(R)$ models stronger than F5, whereas weaker models are difficult to distinguish from GR for this sample of clusters. We show our constraints of the parameter values in Table~\ref{table:fr_pipeline}. Since the $\log_{10}|f_{R0}|$ posterior does not follow a normal distribution, we quote an upper bound rather than the mean and standard deviation. In this case, 68\% of the sampled points have $\log_{10}|f_{R0}|\leq-5.56$. We note that this threshold may depend on the width of the $\log_{10}|f_{R0}|$ prior: for a wider prior (i.e., extending the lower bound of the prior to some value smaller than $-7$ while fixing the upper bound of the prior) and a uniform $\log_{10}|f_{R0}|$ posterior, it is reasonable to expect the 68\% upper bound to be lower. Therefore, it is perhaps more useful to look at the combination of parameter values that give the highest log-likelihood. In this case, the most likely combination has $\log_{10}|f_{R0}|=-6.75$, which is quite close to the lower bound of the prior (although we note that, given the flat posterior distribution of $\log_{10}|f_{R0}|$, the point with $\mathcal{L}_{\rm max}$ might not be much more significant than points with only slightly smaller log-likelihood values). The constraints for the other parameters are in excellent agreement with the fiducial values. Therefore, the results suggest that our pipeline can successfully constrain $f_{R0}$ using cluster samples in a GR universe.

The constraints for the F5 mock are indicated by the blue contours in Fig.~\ref{fig:full_fr_pipeline}. The $\log_{10}|f_{R0}|$ constraints appear to be in good agreement with the fiducial value $-5$, which lies within the 68\% confidence region of the contours. This region only extends down to $\log_{10}|f_{R0}|\approx-6.5$, clearly favouring $f(R)$ gravity over GR. The constraints also appear to rule out models with $\log_{10}|f_{R0}|\gtrsim-4.5$. The median and 68-percentile range of the sampled values is $\log|f_{R0}|=-5.1^{+0.3}_{-1.0}$, while the highest-likelihood parameter combination has $\log_{10}|f_{R0}|=-4.92$. Both of these results are very close to the fiducial value of $-5$. The constraints for the other parameters are again in very reasonable agreement with the fiducial values. This result suggests that our pipeline can clearly identify if the underlying universe model is F5.

Despite this promising agreement, it is interesting to note that the $\log_{10}|f_{R0}|$ posterior distribution has a long tail over the range $-7<\log_{10}|f_{R0}|<-5$. Over this range of points, $\sigma_8$ appears to have value $0.83$-$0.84$ on average, while $\alpha$ and $\beta$ have values $\sim1.75$ and $\sim0.0$ on average (see the blue contours in Fig.~\ref{fig:full_fr_pipeline}. As $\log_{10}|f_{R0}|$ is lowered, the predicted amplitude of the HMF will be reduced. The increased $\sigma_8$ can act against this, as can the lowered $\alpha$, which, as discussed in Sec.~\ref{sec:results:gr_pipeline}, can increase the predicted cluster count for clusters with $0.8M_{500}<6\times10^{14}M_{\odot}$. The latter can also give a scaling relation that more closely matches the F5 result: this is because the scaling relation in F5 is enhanced at lower masses, which may be approximated by the constraint pipeline as a power-law with shallower slope. This degeneracy also comes into play for $\log_{10}|f_{R0}|>-5$, where $\sigma_8$ becomes slightly lower on average and $\alpha$ and $\beta$ become higher. Overall, this reduces the precision of the $\log_{10}|f_{R0}|$ constraint, and is perhaps the reason why the $\log_{10}|f_{R0}|$ posterior peaks at a value that is slightly higher than $-5$. This can also explain why the $\beta$ constraints predict a value $0.3\pm0.4$ that is slightly lower than the fiducial value 0.66. By using tighter priors in $\sigma_8$, $\alpha$ or $\beta$ it may be possible to eliminate this bias (see Sec.~\ref{sec:bias:degeneracies} for a detailed discussion).

We have also tested our pipeline using an F4.5 mock (with $\log_{10}|f_{R0}|=-4.5$), and in Appendix \ref{sec:appendix:F4.5} we show that this model is clearly distinguished from F5.

\section{Potential biases in model constraints}
\label{sec:bias}

In Sec.~\ref{sec:results}, we demonstrated that our framework can give very reasonable constraints of $\log_{10}|f_{R0}|$ for both GR and F5 mocks (Fig.~\ref{fig:full_fr_pipeline}). An important feature of this constraint framework (Fig.~\ref{fig:flow_chart}) is the inclusion of corrections for the effects of $f(R)$ gravity on the internal cluster properties, which are expected to prevent biased constraints. In Sec.~\ref{sec:bias:pipeline}, we will assess potential sources of bias in the constraint pipeline, including an incomplete treatment of the scaling relation. Then, in Sec.~\ref{sec:bias:sample}, we will check the effects of the cluster sample, including selection criteria, on the constraints. Finally, we will demonstrate how the various parameter degeneracies can be prevented by using tighter parameter priors in Sec.~\ref{sec:bias:degeneracies}. 

For all of the figures in this section, we will only show constraints for parameters that are either biased or contribute to parameter degeneracies. Therefore, we exclude the $\log Y_{\star}$ and $\sigma_{\log Y}$ constraints, since these always match the Gaussian priors very closely (e.g., see Figs.~\ref{fig:gr_pipeline} and \ref{fig:full_fr_pipeline}). For similar reasons, we will also exclude $\Omega_{\rm M}$ constraints that have been inferred using the Gaussian prior from Planck 2018.

\subsection{Constraint pipeline}
\label{sec:bias:pipeline}

\subsubsection{Power-law scaling relation}
\label{sec:bias:pipeline:power_law_sr}

\begin{figure}
\centering
\includegraphics[width=\columnwidth]{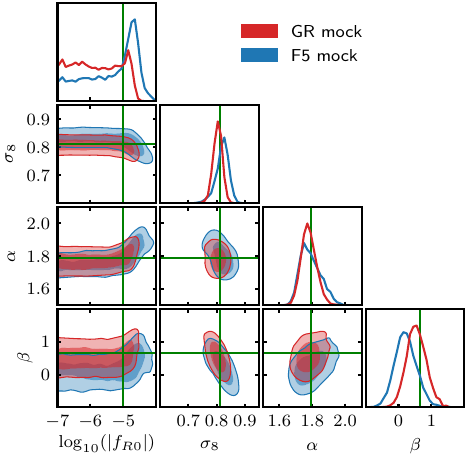}
\caption{[{\it Colour Online}] Parameter constraints generated using the same GR (\textit{red}) and F5 (\textit{blue}) mocks as Fig.~\ref{fig:full_fr_pipeline}, but with a simplified constraint pipeline in which the $f(R)$ effects on the observable-mass scaling relation are neglected.}
\label{fig:gr_sr}
\end{figure}

In Fig.~\ref{fig:gr_sr}, we show constraints inferred using the same GR and F5 mocks as used for Fig.~\ref{fig:full_fr_pipeline}. However, here the $f(R)$ effects on the SZ scaling relation (Eq.~(\ref{eq:planck_ysz})) have been neglected, i.e., a power-law scaling relation without $f(R)$ corrections is used in the (incomplete) $f(R)$ pipeline. 

For the GR mock constraints, shown by the red contours in Fig.~\ref{fig:gr_sr}, the $\log_{10}|f_{R0}|$ posterior appears to be uniformly distributed over the range $-7\leq\log_{10}|f_{R0}|\lesssim-4.5$. This extends beyond the range $-7\leq\log_{10}|f_{R0}|\lesssim-5$ observed using the full pipeline in Fig.~\ref{fig:full_fr_pipeline}, and the range $\log_{10}|f_{R0}|\leq-5.36$ containing 68\% of the sampled points has a higher upper bound than the range $\leq-5.56$ given in Table~\ref{table:fr_pipeline} for the full pipeline. Therefore, even though the GR mock is generated using a power-law scaling relation, it seems that neglecting the $f(R)$ effects on the scaling relation in the pipeline leads to less precise and weaker constraints of $\log_{10}|f_{R0}|$ overall. 

The F5 mock constraints, which are shown by the blue contours, still give a peaked $\log_{10}|f_{R0}|$ posterior distribution. However, there are now a greater proportion of sampled points within the range $-7\leq\log_{10}|f_{R0}|\lesssim-5$. This means that the 68\% confidence contours extend to $\log_{10}|f_{R0}|=-7$, indicating that the pipeline is unable to convincingly rule out GR. There are also a greater number of sampled points with $\log_{10}|f_{R0}|\gtrsim-4.5$; indeed, the highest-likelihood parameter combination has $\log_{10}|f_{R0}|=-4.56$, which is much higher than the fiducial value $-5$ and the value $-4.92$ when using the full pipeline. The median and 68-percentile range is $\log_{10}|f_{R0}|=-5.1^{+0.5}_{-1.2}$, which is less precise than the constraint $\log_{10}|f_{R0}|=-5.1^{+0.3}_{-1.0}$ with the full $f(R)$ pipeline. 

In summary, our constraints for the GR and F5 mocks indicate that assuming a power-law observable-mass scaling relation can lead to imprecise and biased constraints of $f(R)$ gravity. This appears to be linked to parameter degeneracies, where we again observe a lowered $\sigma_8$ and increased $\alpha$ for $\log_{10}|f_{R0}|\gtrsim-5$, and an increased $\sigma_8$ and lowered $\alpha$ and $\beta$ at $\log_{10}|f_{R0}|\lesssim-5$.

\subsubsection{Mass ratio scatter}
\label{sec:bias:pipeline:ratio_scatter}

\begin{figure}
\centering
\includegraphics[width=\columnwidth]{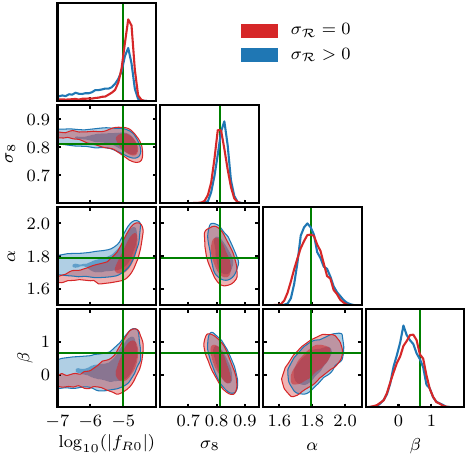}
\caption{[{\it Colour Online}] Parameter constraints generated using our constraint pipeline, where the blue constraints are the same as the F5 mock constraints in Fig.~\ref{fig:full_fr_pipeline} and the red constraints are generated with the scatter of the dynamical mass enhancement set to zero in both the mock and log-likelihood.}
\label{fig:corner_ratio_scatter}
\end{figure}

For our constraints using the F5 mock in Fig.~\ref{fig:full_fr_pipeline}, we included the scatter of the dynamical mass enhancement, given by Eq.~(\ref{eq:scatter_model}), in both the mock and the log-likelihood calculation. We now consider the effect of neglecting this scatter from the mock and the likelihood. The new result is shown by the red contours in Fig.~\ref{fig:corner_ratio_scatter}, along with the previous results in blue. Without this scatter, the observable-mass scaling relation is less scattered overall; as a result, the $f(R)$ constraints are more precise, with 68-percentile range $\log_{10}|f_{R0}|=-4.89^{+0.15}_{-0.35}$ as opposed to $\log_{10}|f_{R0}|=-5.1^{+0.3}_{-1.0}$. In particular, the red 68\% contours do not feature the tail towards low $\log_{10}|f_{R0}|$. These results indicate that excluding the scatter could lead to $f(R)$ constraints with an unrealistically high precision.

\subsection{Cluster sample}
\label{sec:bias:sample}

\subsubsection{Flux threshold}
\label{sec:bias:sample:ysz_cut}

\begin{figure}
\centering
\includegraphics[width=\columnwidth]{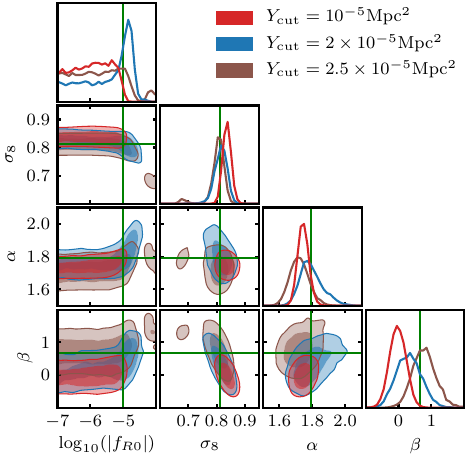}
\caption{[{\it Colour Online}] Parameter constraints generated using our constraint pipeline, using F5 mocks with observational flux thresholds of $10^{-5}{\rm Mpc}^2$ (\textit{red}), $2\times10^{-5}{\rm Mpc}^2$ (\textit{blue}) and $2.5\times10^{-5}{\rm Mpc}^2$ (\textit{brown}).}
\label{fig:corner_ycut}
\end{figure}

In addition to the observational cut $Y_{\rm cut}=1.5\times10^{-5}{\rm Mpc}^2$ which is used in the main results of this paper, we have also generated mocks with cuts $10^{-5}{\rm Mpc^2}$, $2\times10^{-5}{\rm Mpc^2}$ and $2.5\times10^{-5}{\rm Mpc^2}$. From Fig.~\ref{fig:mock_sr}, a cut of $10^{-5}{\rm Mpc}^2$ means that the lowest mass clusters, with $M_{500}\sim10^{14}h^{-1}M_{\odot}$, are included in the sample. In the F5 model, the HMF is more enhanced at these lower halo masses (see Fig.~\ref{fig:hmf_enhancement}), therefore it is expected that using lower-mass objects can give more precise constraints of $\log_{10}|f_{R0}|$. 

In Fig.~\ref{fig:corner_ycut}, we show constraints generated from F5 mocks with these three cuts. For $Y_{\rm cut}=2.5\times10^{-5}{\rm Mpc}^2$, the sampled $\log_{10}|f_{R0}|$ distribution is quite uniform for $-7<\log_{10}|f_{R0}|\lesssim-5$, indicating that this high-mass cluster sample cannot be used to distinguish the F5 model from weaker models, including GR. This is not surprising, given that higher-mass clusters are better-screened in F5, which means that their number count deviates from the GR prediction less strongly (see Fig.~\ref{fig:hmf_enhancement}). On the other hand, the constraints for $Y_{\rm cut}=2\times10^{-5}{\rm Mpc}^2$ clearly favour $\log_{10}|f_{R0}|$ values close to $-5$. However, the 68\% contours still extend to $\log_{10}|f_{R0}|=-7$, which is very close to GR. This is improved upon with $Y_{\rm cut}=1.5\times10^{-5}{\rm Mpc}^2$, which is able to convincingly distinguish the F5 model from GR, as we showed in Fig.~\ref{fig:full_fr_pipeline}. 

In Fig.~\ref{fig:corner_ycut}, we also show the constraints from the F5 mock with $Y_{\rm cut}=10^{-5}{\rm Mpc}^2$. Interestingly, despite containing lower-mass clusters than the other mocks, the sampled $\log_{10}|f_{R0}|$ values are approximately evenly distributed over $-7\lesssim\log_{10}|f_{R0}|\lesssim-5$. One possible reason is that this mock catalogue includes many more low-mass, unscreened, clusters, and the main constraining power comes from different objects than the previous cases. We note that for these constraints, the $\sigma_8$, $\alpha$ and $\beta$ parameters are all biased. As we have already discussed, these parameters can be varied in such a way that the predicted theoretical HMF in GR (i.e., with low $\log_{10}|f_{R0}|$) can match the F5 HMF with the fiducial cosmological parameters. Our results here show that this can cause biased constraints which appear to prefer GR over $f(R)$ gravity even though this is an F5 mock, and this seems to be more relevant for cluster samples that extend to lower masses. As we will show in Sec.~\ref{sec:bias:degeneracies}, these degeneracies can be prevented by using tighter parameter priors.

\begin{figure}
\centering
\includegraphics[width=\columnwidth]{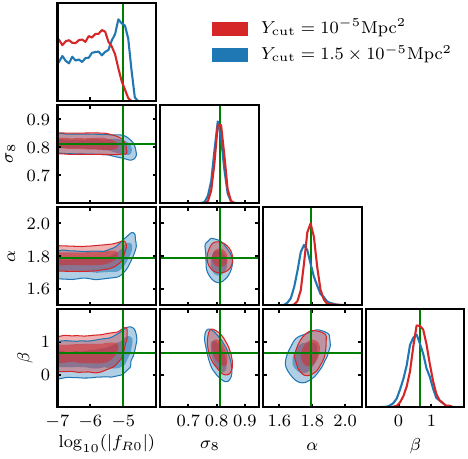}
\caption{[{\it Colour Online}] Parameter constraints generated using our constraint pipeline, using GR mocks with observational flux thresholds $10^{-5}{\rm Mpc}^2$ (\textit{red}) and $1.5\times10^{-5}{\rm Mpc}^2$ (\textit{blue}). The latter is a different realisation (generated in the same way) from the GR mock used in Fig.~\ref{fig:full_fr_pipeline}, and is included to show the potential effects of sample variance on the constraints.}
\label{fig:corner_sample_variance}
\end{figure}

We note that the biased results described above only apply to an F5 fiducial cosmology. The red contours in Fig.~\ref{fig:corner_sample_variance} show the constraints inferred using a GR mock with $Y_{\rm cut}=10^{-5}{\rm Mpc}^2$. These are consistent with GR, with 68\% of the sampled points in the range $\log_{10}|f_{R0}|\leq-5.71$, which is even more precise than the $\log_{10}\leq-5.56$ constraint from Fig.~\ref{fig:full_fr_pipeline}. Meanwhile, the constraints for $\sigma_8$, $\alpha$ and $\beta$ show an excellent match with the fiducial values. Therefore, the bias described above may not be an issue for cluster samples in a GR universe.

\subsubsection{Sample variance}
\label{sec:bias:sample:variance}

In order to check the effect of sample variance on the constraints, we have generated several GR mocks with $Y_{\rm cut}=1.5\times10^{-5}{\rm Mpc}^2$, following the method discussed in Sec.~\ref{sec:methods:mock}. In all cases, the inferred constraints of $\log_{10}|f_{R0}|$ are consistent with GR, with the 68\% constraint contours spanning $-7\leq\log_{10}|f_{R0}|\lesssim-5$ just like the red contours in Fig.~\ref{fig:full_fr_pipeline}. 

However, we have occasionally observed peaks in the $\log_{10}|f_{R0}|$ posterior distribution close to $-5$, which are related to the degeneracies between $\log_{10}|f_{R0}|$, $\sigma_8$, $\alpha$ and $\beta$ mentioned above. An example is shown with the blue contours in Fig.~\ref{fig:corner_sample_variance}. As we have discussed, in the constraints using the F5 mock in Fig.~\ref{fig:full_fr_pipeline}, we can see a `rise' in the $\log_{10}|f_{R0}|$--$\alpha$ contour at $\log_{10}|f_{R0}|>-5$; there is a similar `rise' in the case of the blue contours in Fig.~\ref{fig:corner_sample_variance}. This is because a larger $\alpha$, which means a steeper scaling relation and hence underpredicted cluster number counts, could be compensated by a stronger gravity, so that to the pipeline, the GR mock would appear to be reasonably fitted with an $f(R)$ model with slightly larger $\alpha$. We also see a slight `drop' in the $\log_{10}|f_{R0}|$--$\sigma_8$ contour, where the lowered $\sigma_8$ can again counteract the strengthened gravity. These effects can lead to more points sampled around $\log_{10}|f_{R0}|=-5$, and because even stronger gravity is disfavoured an artificial peak is formed at $-5$. While the peak in $\log_{10}|f_{R0}|$ here is smaller than the peak observed for the F5 mock in Fig.~\ref{fig:full_fr_pipeline}, it is important to be wary that degeneracies can lead to a particular value of $\log_{10}|f_{R0}|$ being favoured even for a GR fiducial cosmology. Like the other sources of bias discussed in this work, this issue can be eliminated by using tighter priors, as we will show in the next section.

\subsection{Tighter priors}
\label{sec:bias:degeneracies}

\begin{figure*}
\centering
\includegraphics[width=0.75\textwidth]{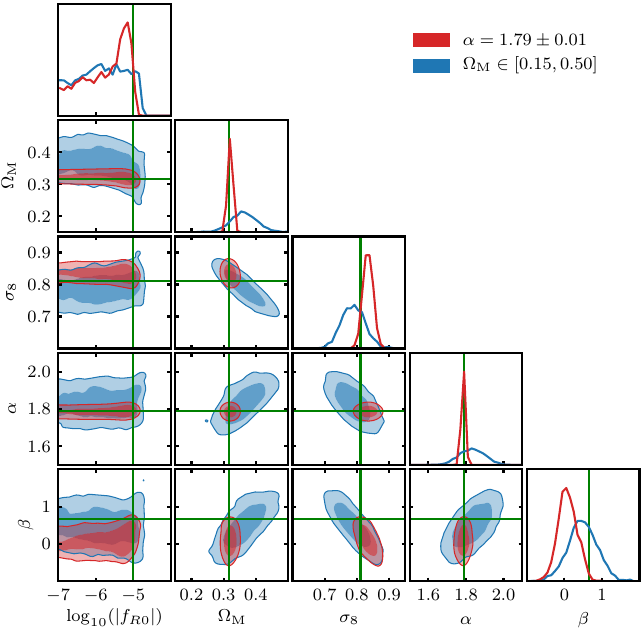}
\caption{[{\it Colour Online}] Parameter constraints generated by our constraint pipeline using: an F5 mock with flux threshold $Y_{\rm cut}=10^{-5}{\rm Mpc}^2$ and a tight Gaussian prior $1.79\pm0.01$ for $\alpha$ (\textit{red}); and an F5 mock with flux threshold $Y_{\rm cut}=1.5\times10^{-5}{\rm Mpc}^2$ and a flat prior $[0.15,0.50]$ for $\Omega_{\rm M}$ (blue).}
\label{fig:corner_tight_priors}
\end{figure*}

\begin{figure*}
\centering
\includegraphics[width=\textwidth]{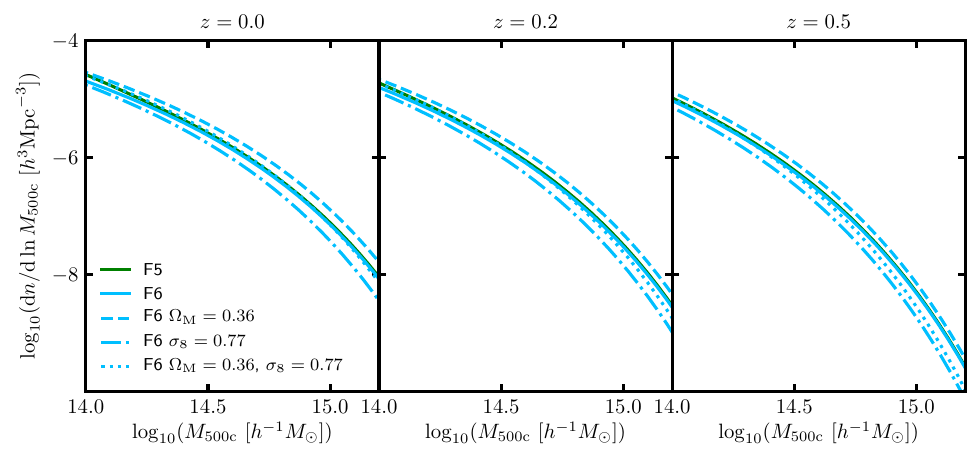}
\caption{[{\it Colour Online}] Predictions of the HMF in F6 (\textit{light blue}) and F5 (\textit{dark green}) at redshifts 0.0, 0.2 and 0.5. We show predictions with the fiducial parameter values $\Omega_{\rm M}=0.3153$ and $\sigma_8=0.8111$ (\textit{solid lines}), an increased $\Omega_{\rm M}$ (\textit{dashed line}), a reduced $\sigma_8$ (\textit{dash-dotted line}) and both an increased $\Omega_{\rm M}$ and reduced $\sigma_8$ (\textit{dotted line}). This figure illustrates not only the well-known degeneracy between $\Omega_{\rm M}$ and $\sigma_8$ in determining the HMF, but also their degeneracy with $f_{R0}$: by tuning the values of these two parameters, an F6 model can closely mimic the HMF of an F5 model; note that the latter degeneracy may be broken by looking at multiple redshifts or by having more precise knowledge of $\Omega_{\rm M}$ and/or $\sigma_8$.}
\label{fig:hmf_degeneracy}
\end{figure*}

For the many of the results discussed in this work, we have observed degeneracies between $\log_{10}|f_{R0}|$, $\Omega_{\rm M}$, $\sigma_8$, $\alpha$ and $\beta$. Together, these parameters can vary such that the theoretical GR HMF is consistent with the F5 mocks, or similarly the theoretical F5 HMF can be made consistent with the GR mocks. For our main results with the $f(R)$ constraint pipeline (Fig.~\ref{fig:full_fr_pipeline}), we have been using a tight Gaussian $\Omega_{\rm M}$ prior from Planck 2018. In Fig.~\ref{fig:corner_tight_priors}, the blue constraints have been generated using the same F5 mock as the blue constraints in Fig.~\ref{fig:full_fr_pipeline}; however, here a flat prior [0.15,0.50] has been adopted for $\Omega_{\rm M}$. This gives rise to the degeneracy between $\Omega_{\rm M}$ and $\sigma_8$ (observed earlier in Fig.~\ref{fig:gr_pipeline}), which leads to a uniform distribution in $\log_{10}|f_{R0}|$. Fig.~\ref{fig:hmf_degeneracy} provides an illustration of this degeneracy: here, the HMF prediction for F6 with increased $\Omega_{\rm M}$ and reduced $\sigma_8$ closely resembles the F5 prediction, particularly at lower masses which dominate the mock cluster samples. By using the tight $\Omega_{\rm M}=0.3153\pm0.0073$ prior for our main results in Sec.~\ref{sec:results:fr_pipeline}, we have prevented this issue. The tight prior on $\Omega_{\rm M}$ can potentially be replaced by combining cluster number counts with other cosmological probes that are sensitive to $\Omega_{\rm M}$, such as the CMB.

We have also shown that there is a degeneracy between $\sigma_8$ and the SZ scaling relation parameters $\alpha$ and $\beta$. Although we have used Gaussian priors for the latter, they can still vary enough to cause biased constraints. In Sec.~\ref{sec:bias:sample:ysz_cut}, we found that this degeneracy caused the $\log_{10}|f_{R0}|$ constraints using the F5 mock with $Y_{\rm cut}=10^{-5}{\rm Mpc}^2$ to resemble GR (see Fig.~\ref{fig:corner_ycut}). In Fig.~\ref{fig:corner_tight_priors}, the red contours show the $\log_{10}|f_{R0}|$ constraints for the same mock, but this time using a tighter $\alpha$ prior of $1.79\pm0.01$. The $\log_{10}|f_{R0}|$ posterior distribution now peaks close to the fiducial value $-5$, though the constraints on $\sigma_8$ and $\beta$ are similarly biased as before. In this case, as in Sec.~\ref{sec:bias:sample:ysz_cut}, the constrained $\beta$ value is lower, which means less time evolution; because the time evolution is normalised at $z=0$, this implies that, for a given cluster mass $M_{500}$, the measured observable $Y$ at $z>0$ is smaller than the true value, and so fewer detectable clusters would be predicted. This is compensated by a larger $\sigma_8$ (actually a similar degeneracy can be observed in the GR case, see the $\sigma_8$--$\beta$ contour in Fig.~\ref{fig:gr_pipeline}), but one side effect is that smaller $\log_{10}|f_{R0}|$ values are more likely to be allowed, leading to a uniform posterior distribution in Fig~\ref{fig:corner_ycut}, which is alleviated in Fig.~\ref{fig:corner_tight_priors} with the tighter prior on $\alpha$ but nevertheless not completely eliminated. Looking at the red contours in the left column of Fig.~\ref{fig:corner_tight_priors}, we can see that at $\log_{10}|f_{R0}|\approx-5$, $\beta$ and $\sigma_8$ both match their correct values, which suggests that if we can tighten the prior on either $\sigma_8$ or $\beta$, the constraint on $\log_{10}|f_{R0}|$ can be further improved. 

Therefore, a conclusion from this discussion is that, with better knowledge of the scaling relation parameters, it is possible to reduce the effect of these degeneracies. However, we note that it may be difficult to constrain the scaling relation parameters with even greater precision. In this case, the degeneracies could be prevented by using a synergy with weak lensing data, which can estimate the cluster mass with higher precision. Even if this data is only available for a subset of the clusters, it can still be incorporated in the log-likelihood \citep[e.g.,][]{SPT:2018njh}.

\section{Summary, Discussion and Conclusions}
\label{sec:conclusions}

Ongoing and upcoming astronomical surveys \citep[e.g.,][]{lsst,erosita,Ade:2018sbj} are expected to generate vast galaxy cluster catalogues that will be many time larger than previous data sets. The abundance of clusters is highly sensitive to the strength of gravity on large scales. Therefore, the new catalogues will enable us to probe a wide variety of MG theories which have been proposed to explain the accelerated expansion of the Universe. This work is the latest of a series that aims to develop a robust general framework for unbiased cluster constraints of gravity. So far, we have studied the effects of the fifth forces in Hu-Sawicki $f(R)$ gravity and the nDGP model on cluster properties, including the dynamical mass, the halo concentration and the observable-mass scaling relations. If these effects are not properly accounted for in cluster tests of gravity, the inferred constraints may be biased.

In this paper, we have combined our models for all the $f(R)$ effects into an MCMC pipeline for constraining the amplitude of the present-day background scalar field, $|f_{R0}|$. We have adopted the model from \citet{Cataneo:2016iav} for the $f(R)$ enhancement of the HMF, and used this, along with our model for the enhancement of the halo concentration, to produce a model-dependent prediction of the cluster number counts (Sec.~\ref{sec:methods:hmf}). We have also used our model for the enhancement of the dynamical mass in $f(R)$ gravity to convert a GR power-law observable-mass scaling relation, which is based on the Planck $Y_{\rm SZ}(M_{500})$ relation \citep{Planck_SZ_cluster}, into a form consistent with $f(R)$ gravity, where the fifth force enhances the relation at sufficiently low masses (Sec.~\ref{sec:methods:scaling_relation}). These models are all incorporated in our log-likelihood (Sec.~\ref{sec:methods:likelihood}), which we have used to infer parameter constraints using a set of mock cluster catalogues (Sec.~\ref{sec:methods:mock}).

Using a combination of GR and F5 mocks, we have shown that our pipeline is able to give reasonable parameter constraints that are consistent with the fiducial cosmology (Figs.~\ref{fig:gr_pipeline} and \ref{fig:full_fr_pipeline}). For the GR mock, the constraints conclusively rule out $f(R)$ models with $\log_{10}|f_{R0}|\gtrsim-5$ and favour values in the range $-7\leq\log_{10}|f_{R0}|\lesssim-5$ where $-7$ is the lowest value considered by our MCMC sampling. Meanwhile, the constraints inferred using the F5 mock favour values close to the fiducial value of $-5$, with 68\% range $-5.1^{+0.3}_{-1.0}$ and a `most likely' value of $-4.92$. We have also shown that the constraints inferred from both mocks can be imprecise and biased if the $f(R)$ enhancement of the scaling relation is not accounted for (Fig.~\ref{fig:gr_sr}). Therefore, this should be properly modelled in future tests of $f(R)$ gravity in order to prevent biased constraints. This will become particularly relevant as cluster catalogues start to enter the galaxy group regime \citep[e.g.,][]{Pillepich:2018sin,2021Univ....7..139L}, where more objects can be unscreened in $f(R)$ gravity.

Throughout this work, the main obstacle to precise and unbiased constraints has stemmed from degeneracies between $f_{R0}$, $\Omega_{\rm M}$, $\sigma_8$ and the scaling relation parameters $\alpha$ and $\beta$, all of which can influence the predicted cluster count. We have shown that the degeneracies can be prevented by using a tighter Gaussian prior for $\Omega_{\rm M}$ and by having better knowledge of the scaling relation parameters (Fig.~\ref{fig:corner_tight_priors}). The latter can potentially be achieved by including lensing data for a subset of the clusters. If wide or flat parameter priors are used, this may give rise to biased constraints of $\log_{10}|f_{R0}|$. For example, we have found that the parameter degeneracies can have a more significant effect for cluster samples that extend to lower masses (Sec.~\ref{sec:bias:sample:ysz_cut}).

In the near future, we plan to further improve this pipeline in a few ways. First, while the HMF model of \citet{Cataneo:2016iav} is accurate, it only covers the redshift range $[0,0.5]$, and we need an extended model that works for a larger redshift range, as well as for wider ranges of other cosmological parameters (not restricted to the $\Omega_{\rm M}$ and $\sigma_8$ parameters as we have focused on here). Calibrating this model for spherical overdensity $\Delta=500$ would also mean that conversions between halo mass definitions would no longer be required. Second, we plan to run larger hydrodynamical simulations than those used in \citet{Mitchell:2020fnj}, to further study and calibrate the various cluster scaling relations (not limited to $Y_{\rm SZ}$) in this gravity model. Third, the MCMC pipeline will be extended so that we can include independent cluster data, such as weak lensing, in the model constraint. Once these tasks are completed, we can use this pipeline to constrain $f(R)$ gravity using observations. It is also straightforward to extend our framework to other gravity models; we have already started to do this for the nDGP model \citep{Mitchell:2021aex}, where we have provided fitting formulae for the HMF and concentration, and studied the cluster scaling relations, in this model.

\section*{Acknowledgements}

We thank Matteo Cataneo for useful discussions on the halo mass function fitting model and sharing his code for comparisons. MAM is supported by a PhD Studentship with the Durham Centre for Doctoral Training in Data Intensive Science, funded by the UK Science and Technology Facilities Council (STFC, ST/P006744/1) and Durham University. CA and BL are supported by the European Research Council via grant ERC-StG-716532-PUNCA. BL is additionally supported by STFC Consolidated Grants ST/T000244/1 and ST/P000541/1. This work used the DiRAC@Durham facility managed by the Institute for Computational Cosmology on behalf of the STFC DiRAC HPC Facility (\url{www.dirac.ac.uk}). The equipment was funded by BEIS capital funding via STFC capital grants ST/K00042X/1, ST/P002293/1, ST/R002371/1 and ST/S002502/1, Durham University and STFC operations grant ST/R000832/1. DiRAC is part of the National e-Infrastructure.

\section*{Data availability}

The simulation data used in this paper may be available upon request to the corresponding author.




\bibliographystyle{mnras}
\bibliography{references} 



\appendix

\section{Modelling the dynamical mass scatter}
\label{sec:appendix:mdyn_scatter}

In Fig.~\ref{fig:rms_scatter}, the data points show the binned mass ratio scatter as a function of the rescaled logarithmic mass,  $\log_{10}(M_{500}M_{\odot}^{-1}h)-p_2=\log_{10}(M_{500}/10^{p_2})$. To generate this, we have evaluated the difference between the actual dynamical mass enhancement and the value predicted by Eq.~(\ref{eq:mdyn_enhancement}) for each halo, and measured the root-mean-square difference within the same mass bins as used to fit Eq.~(\ref{eq:mdyn_enhancement}) in \citet{Mitchell:2018qrg}. We have modelled this data using a 6-parameter fitting formula which is made up of two parts. A skewed normal distribution is used to capture the shape of the peak: this includes parameters for the normalisation $\lambda_{\rm s}$, the position $\mu_{\rm s}$ and width $\sigma_{\rm s}$ with respect to the $x=\log_{10}(M_{500}/10^{p_2})$ axis, and a parameter $\alpha$ quantifying the skewness. On its own, this distribution would fall to zero at both low and high $x$; however, we see from Fig.~\ref{fig:rms_scatter} that the scatter is slightly greater on average at high $x$ than at low $x$. To account for this, we add on a $\tanh$ function with two parameters: an amplitude $\lambda_{\rm t}$ and a shift $y_{\rm t}$ along the vertical axis. Our full model is then given by:
\begin{equation}
\sigma_{\mathcal{R}} = \frac{\lambda_{\rm s}}{\sigma_{\rm s}}\phi(x')\left[1+\rm{erf}\left(\frac{\alpha x'}{\sqrt[]{2}}\right)\right] + \left(\lambda_{\rm t}\tanh(x)+y_{\rm t}\right),
\label{eq:scatter_model}
\end{equation}
where $x'=(x-\mu_{\rm s})/\sigma_{\rm s}$. $\phi(x')$ represents the normal distribution:
\begin{equation}
\phi(x') = \frac{1}{\sqrt[]{2\pi}}\exp\left(-\frac{x'^2}{2}\right),
\label{eq:normal_dist}
\end{equation}
and ${\rm erf}(x')$ is the error function:
\begin{equation}
{\rm erf}(x') = \frac{2}{\sqrt[]{\pi}}\int_0^{x'}e^{-t^2}{\rm d}t.
\label{eq:err_func}
\end{equation}
Since we have many more data points at higher masses than at lower masses in Fig.~\ref{fig:rms_scatter}, we have used a weighted least squares approach which ensures that different parts of the $\log_{10}(M_{500}/10^{p_2})$ range have an equal contribution to the fitting of Eq.~(\ref{eq:scatter_model}). To do this, we have split the rescaled mass range into 10 equal-width bins and counted the number, $N_i$, of data points within each bin $i$. In the least squares fitting, each data point is then weighted by $1/N_i$. This means that points found at lower masses, where there are fewer data points, are each given a greater weight than points found at higher masses. The resulting best-fit parameter values are: $\lambda_{\rm s}=0.0532\pm0.0008$, $\sigma_{\rm s}=0.58\pm0.03$, $\mu_{\rm s}=-0.35\pm0.03$, $\alpha=1.09\pm0.18$, $\lambda_{\rm t}=0.0012\pm0.0003$ and $y_{\rm t}=0.0019\pm0.0002$.

\section{Mass conversions}
\label{sec:appendix:mass_conversions}

The following formula can be used to convert the HMF from mass definition $M_{\Delta}$ to a new definition $M_{\Delta'}$:
\begin{equation}
    n'(M_{\Delta'}) = n(M_{\Delta}(M_{\Delta'}))\left(\frac{{\rm d}\ln M_{\Delta'}}{{\rm d} \ln M_{\Delta}}\right)^{-1},
    \label{eq:hmf_conversion}
\end{equation}
where $n'$ is the HMF in the new mass definition and $n$ is the HMF in the old definition. This requires a relation between the mass definitions. For this, we use the following \citep{Hu:2002we}:
\begin{equation}
    \frac{M_{\Delta}}{M_{200}} = \frac{\Delta}{200}\left(\frac{c_{\Delta}}{c_{200}}\right)^3,
    \label{eq:mass_conversion}
\end{equation}
where $c_{\Delta}$ is the concentration with respect to generic overdensity $\Delta$. The latter can be computed from $c_{200}$ using:
\begin{equation}
    \frac{1}{c_{\Delta}} = x\left[f_{\Delta}=\frac{\Delta}{200}f\left(\frac{1}{c_{200}}\right)\right],
    \label{eq:c_delta}
\end{equation}
where the function $f(x)$ is given by:
\begin{equation}
    f(x) = x^3\left[\ln(1+x^{-1}) - (1+x)^{-1}\right].
    \label{eq:f_of_x}
\end{equation}
Eq.~(\ref{eq:c_delta}) is computed using the inverse of this function. \citet{Hu:2002we} provide an analytical formula which can accurately solve this:
\begin{equation}
    x(f) = \left[a_1f^{2p} + \left(\frac{3}{4}\right)^2\right]^{-\frac{1}{2}} + 2f,
    \label{eq:x_of_f}
\end{equation}
where $p = a_2 + a_3\ln f + a_4(\ln f)^2$ and the parameters have values $a_1=0.5116$, $a_2=-0.4283$, $a_3=-3.13\times10^{-3}$ and $a_4=-3.52\times10^{-5}$. The authors state that this formula has $\sim0.3\%$ accuracy for galaxy and cluster scales.

\section{\boldmath Test of the constraint pipeline on a stronger \texorpdfstring{$\lowercase{f}(R)$}{f(R)} model}
\label{sec:appendix:F4.5}

\begin{figure*}
\centering
\includegraphics[width=\textwidth]{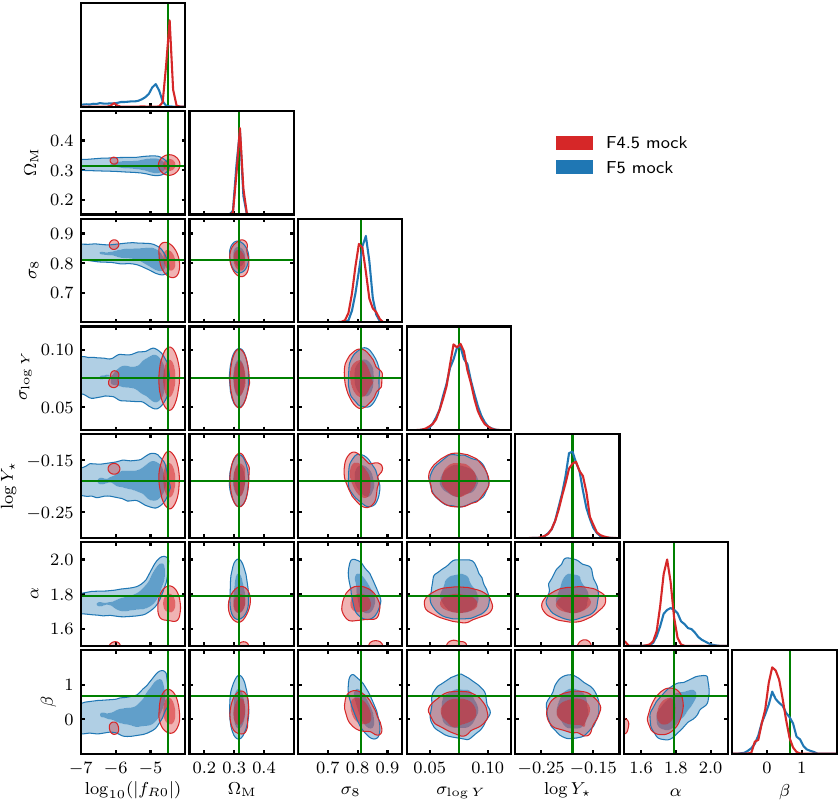}
\caption{[{\it Colour Online}] Parameter constraints obtained by applying our full $f(R)$ pipeline, as detailed in Sec.~\ref{sec:methods}, to F4.5 (\textit{red}) and F5 (\textit{blue}) mocks, with observational flux threshold $Y_{\rm cut}=1.5\times10^{-5}{\rm Mpc}^2$. The dark and light regions of the contours represent 68\% and 95\% confidences, respectively. The marginalised distributions of the sampled parameter values are shown in the top panels of each column. The fiducial cosmological parameter values of the F4.5 mock are indicated by the green lines.}
\label{fig:F4.5_constraints}
\end{figure*}

For the main results of this work, we have tested our constraint pipeline using GR and F5 mocks. For the F5 mock (cf.~Fig.~\ref{fig:full_fr_pipeline}), our pipeline produces a marginalised distribution of $\log_{10}|f_{R0}|$ which peaks close to $-5$, but features a long tail extending to $-7$, which is the lowest value of $\log_{10}|f_{R0}|$ considered in this work. As discussed in Sec.~\ref{sec:results:fr_pipeline}, this can be explained by parameter degeneracies, which can make it more difficult to fully distinguish this model from GR.

To check whether our pipeline can successfully distinguish stronger $f(R)$ models than F5, and whether such models suffer from the same degeneracies, we show, in Fig.~\ref{fig:F4.5_constraints}, constraints obtained using an F4.5 ($\log_{10}|f_{R0}|=-4.5$) mock along with the F5 results from Fig.~\ref{fig:full_fr_pipeline}. The F4.5 constraint features smaller contours and a tight peak at $\log_{10}|f_{R0}|\approx-4.5$ which does not feature long tails towards lower or higher values of $\log_{10}|f_{R0}|$. The median and 68\% range is given by $-4.47^{+0.06}_{-0.07}$, which is in excellent agreement with the fiducial value of $-4.50$. This indicates that our pipeline can clearly distinguish different values of $|f_{R0}|$ and it provides further evidence that it can distinguish $f(R)$ models from GR in an unbiased manner.


\bsp	
\label{lastpage}
\end{document}